\documentclass[fleqn,usenatbib]{mnras}

\usepackage{newtxtext,newtxmath}

\usepackage[T1]{fontenc}
\usepackage{ae,aecompl}

\usepackage{graphicx}	
\usepackage{amsmath}	
\usepackage{amssymb}	
\usepackage{times}

\newcommand{\gaia}		{\emph{Gaia}}

\title[Galactic stellar halo mass]{The total stellar halo mass of the Milky Way}

\author[A. J. Deason et al.]{
Alis J. Deason$^{1}$\thanks{E-mail: alis.j.deason@durham.ac.uk},
Vasily Belokurov$^{2}$, Jason L. Sanders$^{2}$
\\
$^{1}$Institute for Computational Cosmology, Department of Physics, University of Durham, South Road, Durham DH1 3LE, UK\\
$^{2}$Institute of Astronomy, Madingley Road, Cambridge CB3 0HA\\
}

\date{Accepted XXX. Received YYY; in original form ZZZ}

\pubyear{2019}

\begin{document}
\label{firstpage}
\pagerange{\pageref{firstpage}--\pageref{lastpage}}
\maketitle

\begin{abstract}
  We measure the total stellar halo luminosity using red giant branch (RGB) stars selected from \gaia\ data release 2. Using slices in magnitude, colour and location on the sky, we decompose RGB stars belonging to the disc and halo by fitting 2-dimensional Gaussians to the Galactic proper motion distributions. The number counts of RGB stars are converted to total stellar halo luminosity using a suite of isochrones weighted by age and metallicity, and by applying a volume correction based on the stellar halo density profile. Our method is tested and calibrated using \textit{Galaxia} and N-body models.  We find a total luminosity (out to 100 kpc) of $L_{\rm halo} = 7.9 \pm 2.0 \times 10^8L_\odot$ excluding Sgr, and $L_{\rm halo} = 9.4 \pm 2.4 \times 10^8L_\odot$ including Sgr. These values are appropriate for our adopted stellar halo density profile and metallicity distribution, but additional systematics related to these assumptions are quantified and discussed. Assuming a stellar mass-to-light ratio appropriate for a Kroupa initial mass function ($M^\star/L = 1.5$), we estimate a stellar halo mass of $M^\star_{\rm halo} = 1.4 \pm 0.4\times 10^9 M_\odot$. This mass is larger than previous estimates in the literature, but is in good agreement with the emerging picture that the (inner) stellar halo is dominated by one massive dwarf progenitor. Finally, we argue that the combination of a $\sim 10^9M_\odot$ mass and an average metallicity of $\langle \mathrm{[Fe/H]} \rangle \sim -1.5$ for the Galactic halo points to an ancient ($\sim 10$ Gyr) merger event.
\end{abstract}


\section{Introduction}
The halo of our Galaxy is littered with the stellar debris of
destroyed dwarf galaxies. This trash-can of material extends out to
several hundred kiloparsecs, and gives important insight into the
assembly history of the Milky Way and its dark matter
potential. Moreover, the remains of the destroyed dwarfs can tell us
about the properties of the lowest mass galaxies in the Universe.

The content, size, extent, and kinematics of the stellar halo has been
studied extensively over the past few decades (see reviews by
\citealt{helmi08, belokurov13}). In particular, the number counts of
old and relatively metal-poor stars have revealed that the density
profile of the stellar halo approximately follows a power-law with index
$\sim -2.5$ within 20 kpc, and then falls-off more rapidly thereafter,
with power-law index $\sim -4.0$ \citep[e.g.][]{watkins09, sesar10,
  deason11, faccioli14, piladiez15}. Note, however, that the form of the density profile at larger distances ($> 40-50$ kpc) is still highly uncertain \citep[e.g.][]{deason14, xue15, slater16, hernitschek18}. The change in density at $\sim
20$ kpc profile signifies a transition between the ``inner'' and
``outer'' halo. \cite{deason13} argued that this broken profile is
caused by the accretion of a massive dwarf galaxy at early times. In
this scenario, the break radius represents the last apocentre of the
accreted dwarf, and beyond this furthest point of the orbit, the
contribution of the debris from this massive dwarf is significantly
diminished. Thus, this picture suggests that the inner stellar halo is
dominated by one massive accretion event, while the outer halo is a
dusting of several (lower-mass) destroyed dwarfs.
\begin{figure*}
    \begin{minipage}{\linewidth}
        \centering
        \includegraphics[width=0.99\textwidth,angle=0]{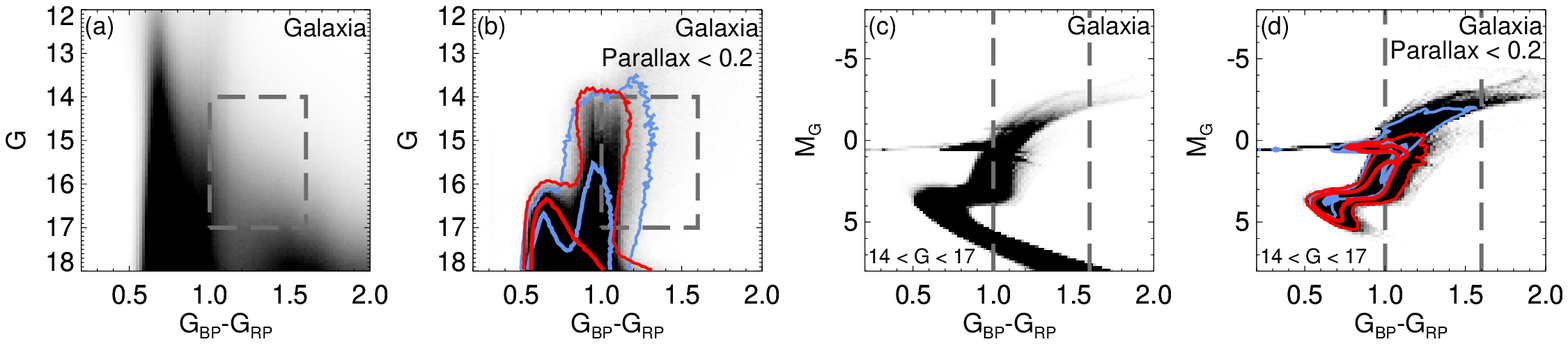}
    \end{minipage}
    \begin{minipage}{\linewidth}
        \centering
        \includegraphics[width=0.99\textwidth,angle=0]{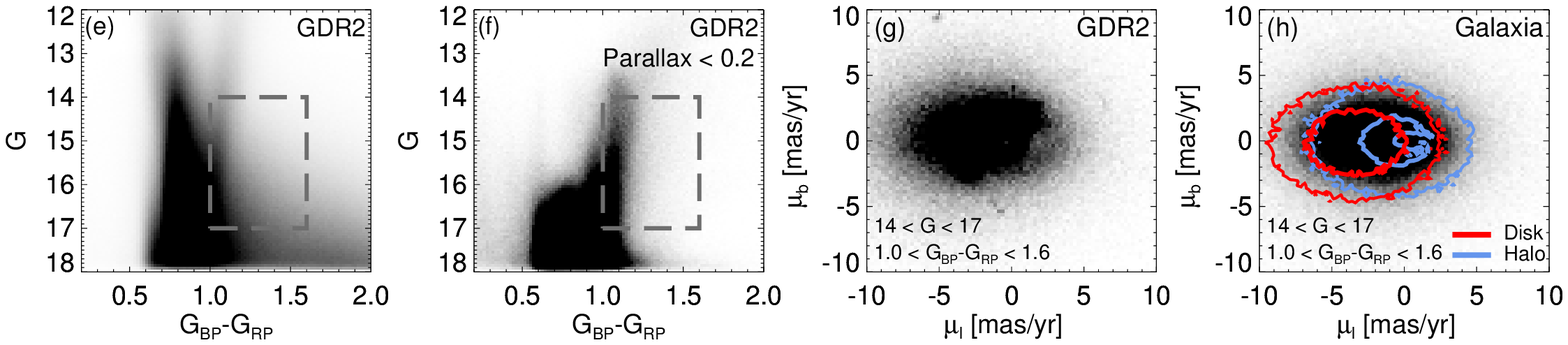}
    \end{minipage}
        \caption[]{Colour magnitude diagrams (CMDs) and proper motion distributions for the \textit{Galaxia} models and GDR2. In all panels, only stars with high latitude ($|b| > 30^\circ$) are shown. \textit{Panel (a):} Apparent magnitude vs. colour for stars in \textit{Galaxia}. Here, the halo component is from an N-body model (Halo-7, see main text). The dashed lines indicate the colour range used in this work to select red giant branch stars. Photometric and astrometric errors applicable to \gaia\ data release 2 have been applied to the model. \textit{Panel (b):} Apparent magnitude vs. colour for stars in \textit{Galaxia} with small parallax (parallax $< 0.2$). This cut removes nearby dwarfs. The red and blue contours indicate the disc and halo stars, respectively. \textit{Panel (c):} Absolute magnitude vs. colour for stars in \textit{Galaxia}. Here, stars with apparent magnitudes $14 < G < 17$ are shown. \textit{Panel (d):}  Absolute magnitude vs. colour for stars in \textit{Galaxia} with small parallax. \textit{Panel (e):} Apparent magnitude vs. colour for stars in GDR2. Note stars in close proximity to the Magellanic Clouds have been removed. \textit{Panel (f):} Apparent magnitude vs. colour for stars in GDR2 with small parallax. \textit{Panel (g):} Proper motions of stars in GDR2 in Galactic coordinates. Here, we only consider stars with parallax $<0.2$, $1.0 < G_{\rm BP} - G_{\rm RP} < 1.6$ and $14 < G < 17$. \textit{Panel (h):} Proper motions in \textit{Galaxia} (with same selection as GDR2). The disc and halo components have distinct, but overlapping, proper motion distributions. These sequences vary across the sky (see Fig. \ref{fig:pm_disk_halo}).}
          \label{fig:cmd}
\end{figure*}

The arrival of the \gaia\ \citep{gaia_miss} data releases
\citep{gaia_dr1, gaia_dr2}, which provide 6-dimensional phase-space
measurements for thousands of local halo stars, and proper motion
measurements for hundreds of thousands of halo stars, reinvigorated
our ideas about the structure of the halo, and confirmed the insight
we gained from the halo number counts. In particular,
\cite{belokurov18}, \cite{haywood18} and \cite{helmi18} used a
combination of kinematical and chemical data from \gaia, SDSS and
APOGEE to find that the inner halo is indeed dominated by one massive
accretion event, which occurred $> 8$ Gyr ago. This significant
event in the history of the Galaxy has been dubbed the \gaia\--Sausage
(aptly named due to it's highly eccentric orbit) or
\gaia\--Enceladus. Follow-up studies have added extra weight to the
growing consensus that the \gaia\--Sausage rules the (inner) halo: for
example, \cite{deason18} and \cite{lancaster19} used the kinematics of
distant halo stars to dynamically show the transition at $\sim 20$ kpc
between the ``Sausage'' dominated regime and the outer halo, and
\cite{myeong18}, \citet{Sequoia} and \citet{Massari2019} used the
dynamics of the globular cluster population in action space to show
that many are likely related to the \gaia\--Sausage, as expected if
this is a massive merger event.
\begin{figure}
  \includegraphics[width=8.5cm, angle=0]{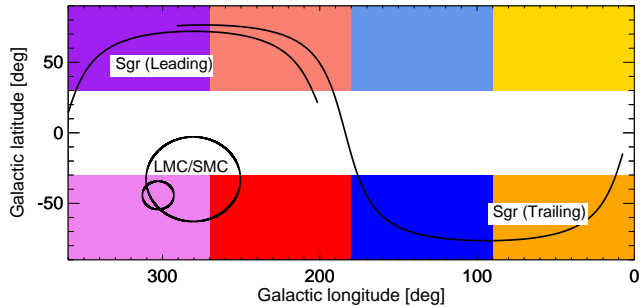}
  \caption{Slices in Galactic longitude and latitude used to fit the
    disc/halo components. Each bin is fitted separately. The colour
    coding indicated is adopted throughout the paper. The Sgr leading
    and trailing arms are shown. When Sgr is excluded, stars lying
    within 12 deg of these tracks are omitted. Stars in close
    proximity to the LMC and/or SMC are excluded in our analysis.}
\label{fig:area}
\end{figure}
\begin{figure*}
        \centering
        \includegraphics[width=0.99\textwidth,angle=0]{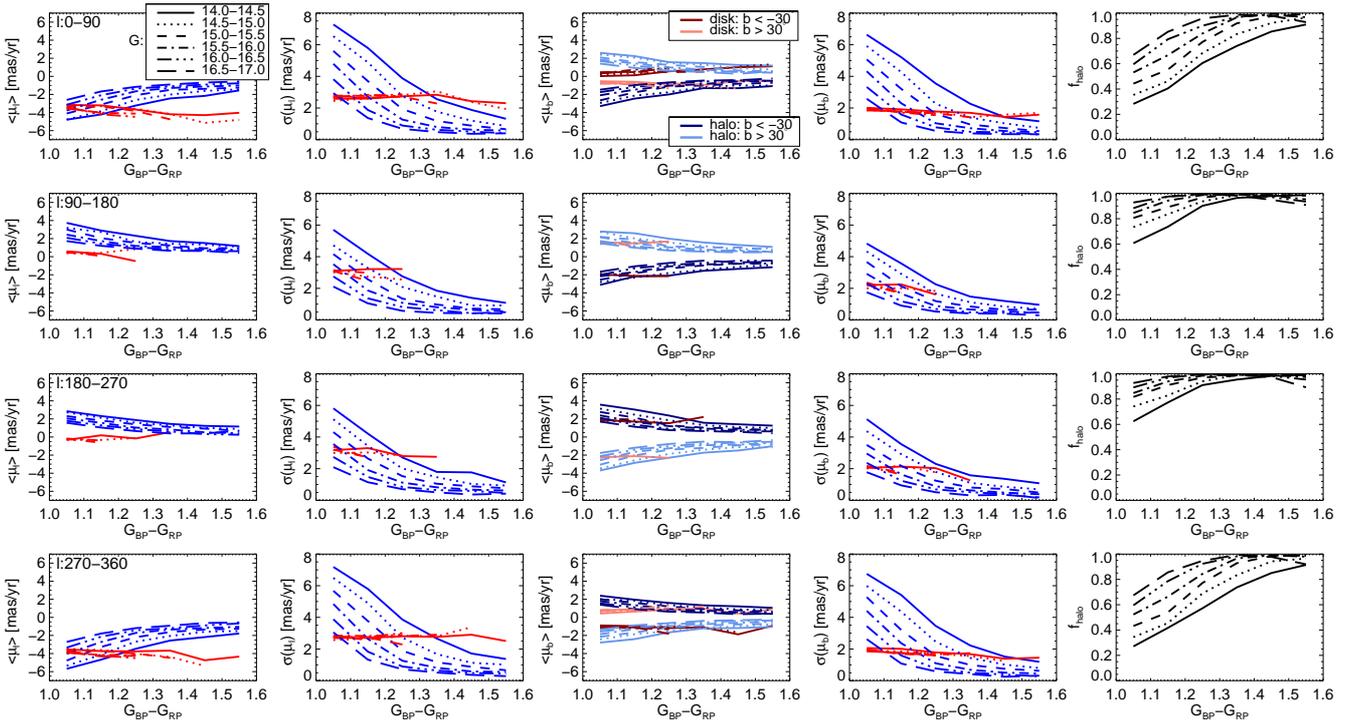}
        \caption{The mean (first and third panels) and dispersion
          (second and fourth panels) of the \textit{Galaxia} model proper
          motions in Galactic coordinates as a function of $G_{\rm BP}
          - G_{\rm RP}$ colour. Blue and red lines indicates the halo
          and disc components, respectively. Different magnitude bins
          are shown with different linestyles, and each row shows a
          different bin in Galactic longitude. The sequences are very
          similar for bins above and below the Galactic plane, except
          for $\langle \mu_b \rangle$ (third column), which we
          indicate with different shades of blue and red. The last
          panel on the right indicates the fraction of halo stars as a
          function of colour.}
          \label{fig:pm_disk_halo}
\end{figure*}

As mentioned above, the density profile of halo stars has proved an
invaluable measure to constrain the Galaxy's assembly
history. However, the normalisation of the density profile, and hence
the total stellar halo mass, has proven to be more complicated to
measure. This is mainly because the tracers we often use to map the
halo star distribution out to large distances, i.e. the RR Lyrae and
blue horizontal branch stars, are difficult to relate to the
\textit{overall} halo population. This is because the exact
broad-brand color (and hence temperature) distribution of the helium
burning stars depends on additional ``hidden'' parameters \citep[see
  e.g.][]{Gratton2010}.

Moreover, it is difficult to provide a robust normalisation when using
surveys that have non-trivial selection functions and/or are limited
in their spatial extent. Most measures of the total stellar halo
density are limited to local halo star samples, and a wide range of
density normalisations have been quoted in the literature: $\rho_0 =
3.0-15.0 \times 10^{-5} M_\odot/\mathrm{pc^3}$ \citep[e.g.][]{morrison93,
  fuchs98, gould98, digby03, juric08, dejong10}. Many of these measures were estimated
before the density profile out to large distances was known, and hence
relating the local stellar density to the total stellar halo mass is
non-trivial. More recently, \cite{bell08} estimated the total stellar
mass using main sequence turn-off stars in SDSS, and \cite{deason11}
used counts of blue horizontal branch stars in SDSS. Both these
studies favour relatively low stellar halo masses $M^\star_{\rm halo}
\sim 3-4 \times 10^8M_\odot$, but there is sizeable uncertainty
relating these tracer populations to the overall stellar halo (see
above). In addition, these measures do not include the few $\sim 10^8
M_\odot$ substructures in the halo, which also contribute to the mass,
so the total mass, based on the \cite{bell08} and \cite{deason11}
estimates for the ``field" halo, is in the range $M^\star_{\rm halo}
\sim 4-7 \times 10^8 M_\odot$ (cf. \citealt{bland-hawthorn16}).

Currently, different estimates of the Galactic stellar halo mass vary 
by a factor of 2, but, more worryingly, the uncertainty of
these estimates is not robustly quantified. Perhaps more puzzling is
that the recent deluge of evidence for a massive accretion event
dominating the stellar halo, appears at odds with the rather low value
of total stellar halo mass quoted in the literature. Rectifying this
apparent conundrum is crucial in order to place the Milky Way in
the cosmological context with other, similar mass galaxies. Both
simulations and observations show that at fixed galaxy (or halo) mass
the range of stellar halo masses is large, reflecting a wide diversity
of assembly histories \citep[e.g.][]{pillepich14, merritt16,
  harmsen17,elias18, canas19, monachesi19}. Moreover, work by \cite{deason16}
and \cite{dsouza18} show that the stellar halo mass is critically
linked to the most massive dwarf progenitor of the halo. Thus, in
order to reconcile several global properties of the Milky Way halo
(e.g. density profile, metallicity) with the currently favoured
assembly history scenario, it is imperative that we procure a robust
total stellar halo mass, complete with a well-defined uncertainty.

In this paper, we utilise the exquisite data from \gaia\ to estimate
the total stellar halo luminosity and mass using red giant branch
(RGB) stars. Compared to previous work, we take advantage of the full
sky coverage of the \gaia\ survey, and use the proper motion
distributions to decompose disc and halo populations. In Section
\ref{sec:rgb} we describe the selection of RGB stars, and introduce
the models that we use to guide and calibrate our analysis. Number
counts of halo stars are estimated in bins of colour, magnitude and
area on the sky, and our process for decomposing the disc and halo
populations is described in Section \ref{sec:decomp}. In Section
\ref{sec:lumf} we determine the normalisation per halo tracer from
stellar population models, and volume correct the number counts in
order to estimate the total stellar halo luminosity. We also quantify
how well this procedure performs on N-body stellar halo models. We
discuss our resulting stellar halo mass in Section \ref{sec:disc}, and
summarise our main conclusions in Section \ref{sec:conc}.

\section{Halo red giant branch stars}
\label{sec:rgb}

\begin{figure*}
  \begin{minipage}{0.33\linewidth}
        \centering
        \includegraphics[width=5.7cm,angle=0]{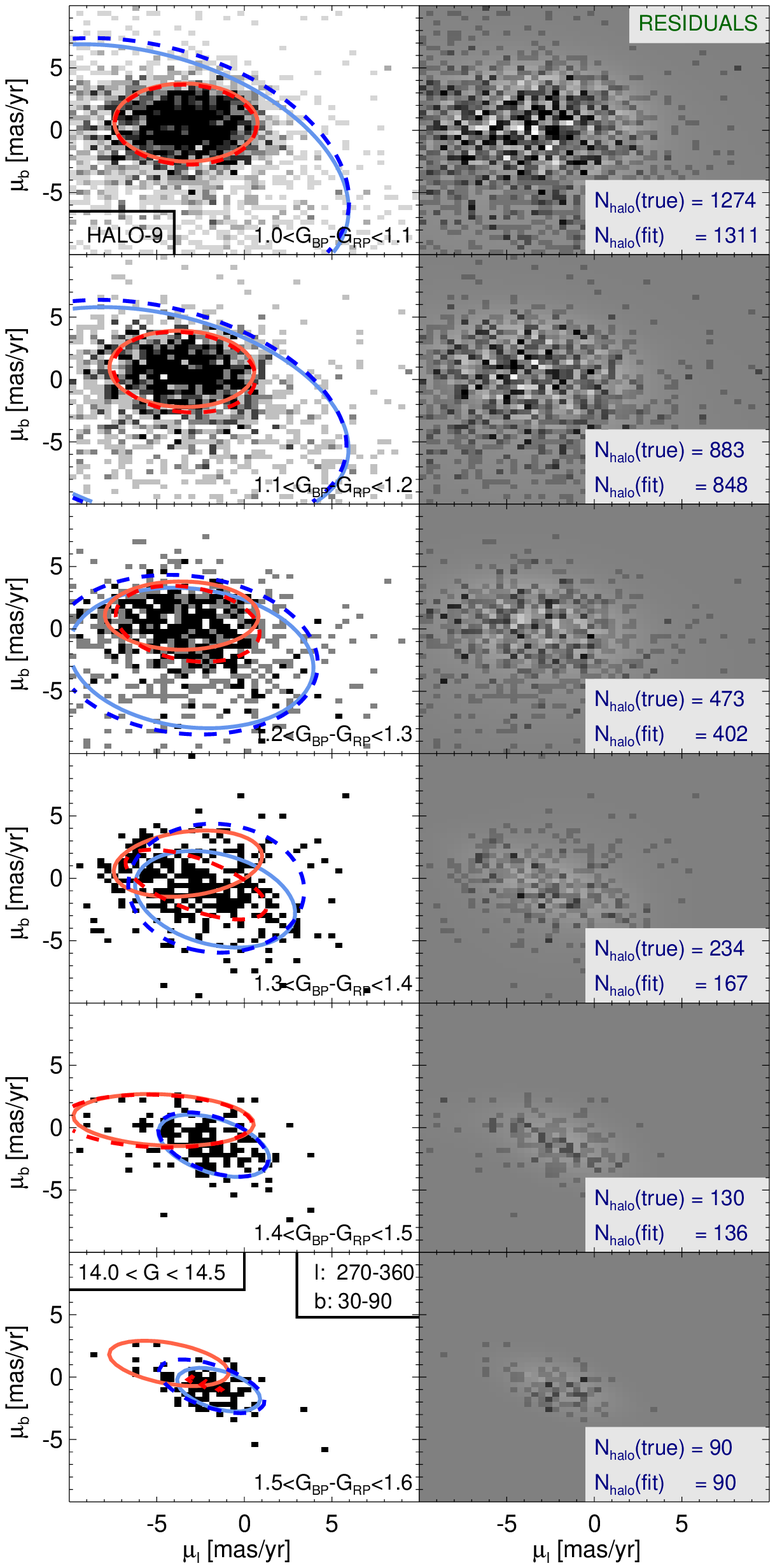}
  \end{minipage}
  \begin{minipage}{0.33\linewidth}
        \centering
        \includegraphics[width=5.7cm,angle=0]{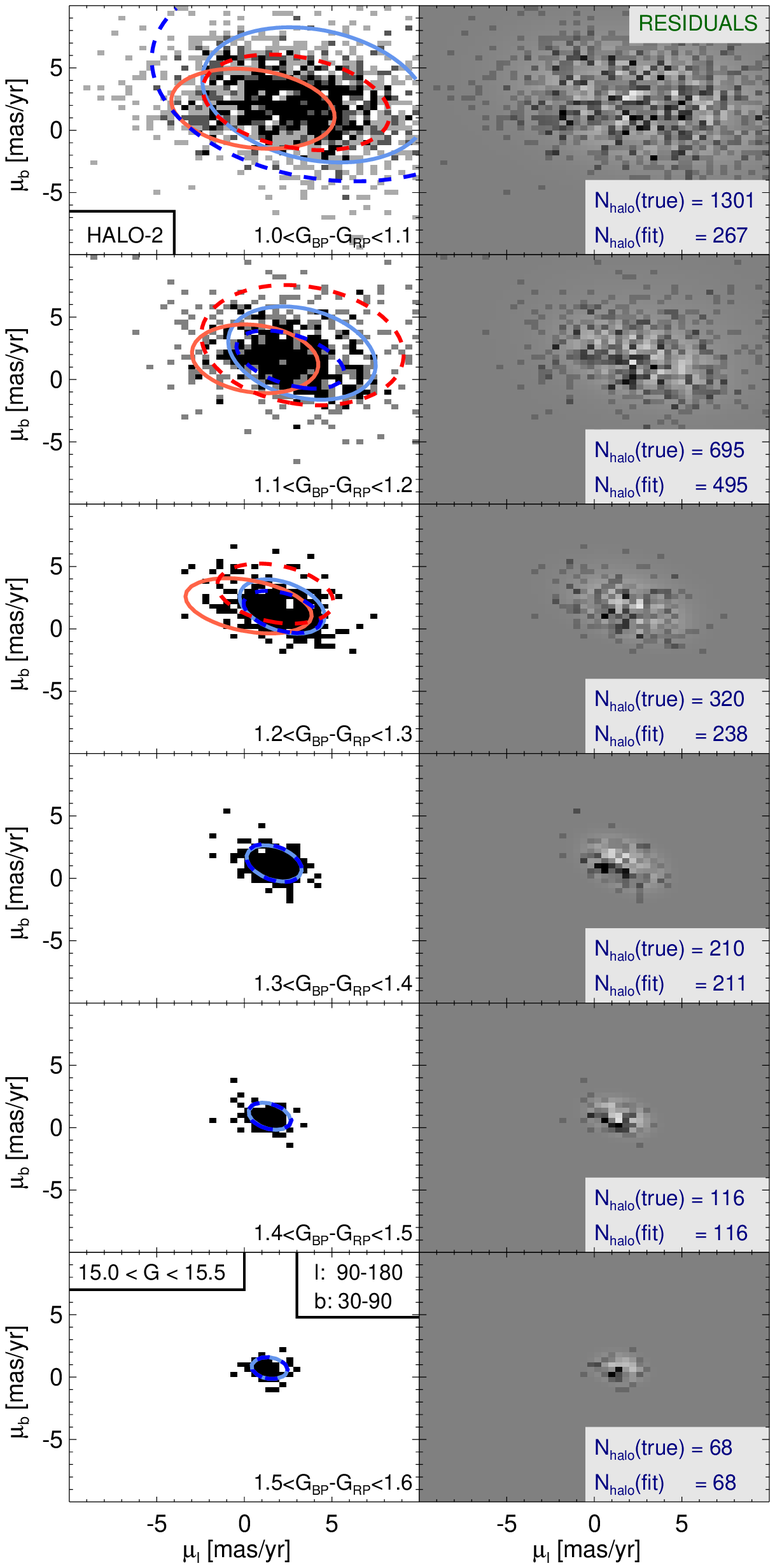}
  \end{minipage}
  \begin{minipage}{0.33\linewidth}
        \centering
        \includegraphics[width=5.7cm,angle=0]{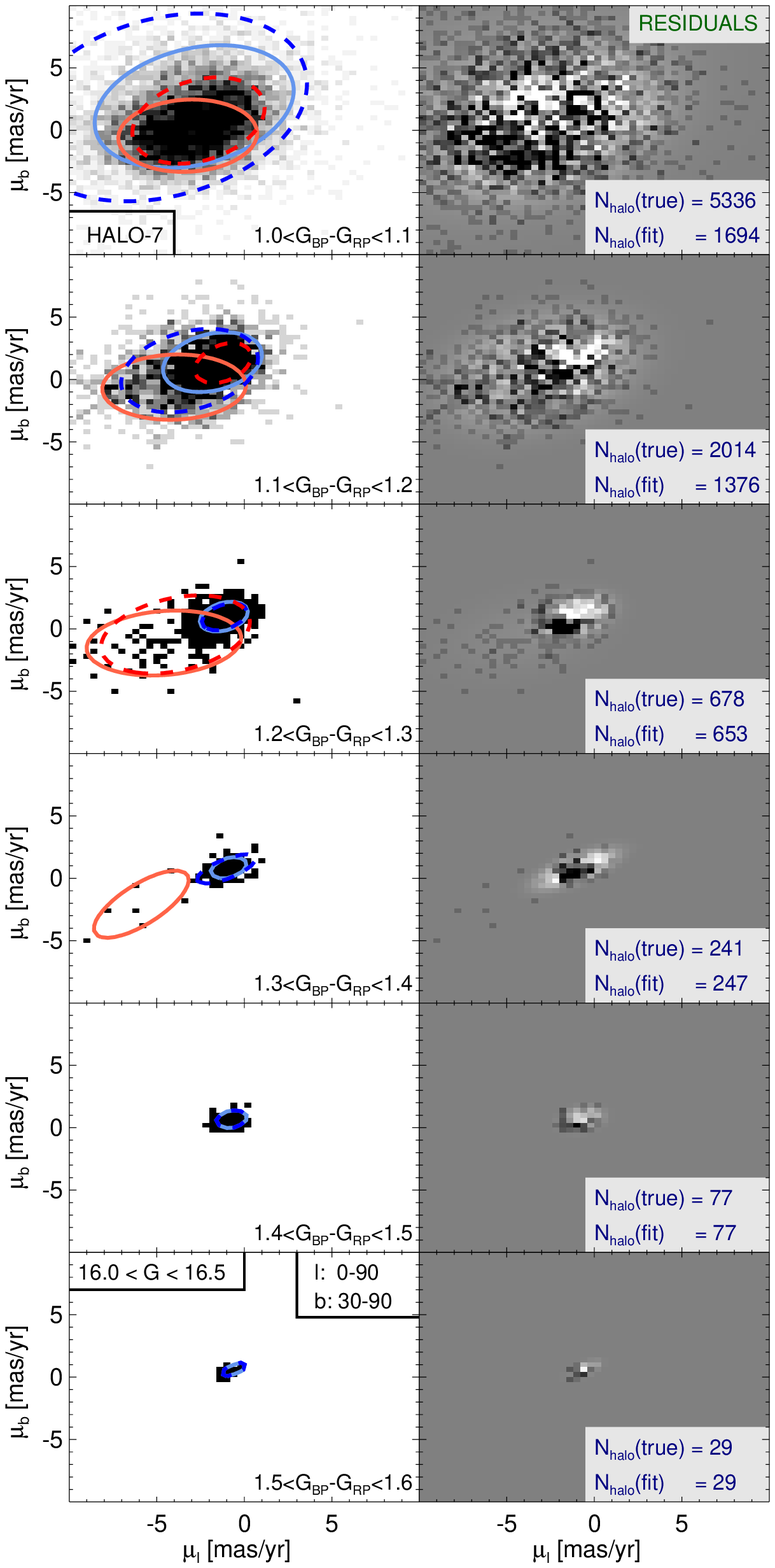}
  \end{minipage}
        \caption{\textit{Left columns:} Extreme deconvolution (XD)
          fits to the Galaxia proper motion distributions in bins of
          $G_{\rm BP} - G_{\rm RP}$ colour.  The solid red and blue
          lines show the true disc and halo distributions, and the
          dashed lines show the XD fits. Here, we use the true Galaxia
          model values to initialize the XD fit. \textit{Right
            columns:} The residuals from the fit. Here, the pixel size
          is 0.4 mas/yr, and the shading saturates at an excess of
          $\Delta = \pm 5$. The true and estimated number of halo
          stars is given in the bottom right hand corner. Three
          examples are shown for different magnitude ranges and bins
          on the sky.}
          \label{fig:resid}
\end{figure*}

Our aim is to use counts of halo red giant branch (RGB) stars to
estimate the total stellar halo luminosity. RGB stars are ideal tracers for
this purpose as they are intrinsically bright, relatively numerous,
and are present at all ages and metallicities. Moreover, we are able
to cleanly select RGB stars using \gaia\ data (see below). In order to
guide us through the stellar halo selection and luminosity estimate,
we make use of ``toy'' models of the Galaxy, which are tailored
towards the \gaia\ data release 2 (GDR2) astrometry and photometry.

\subsection{Galaxia and N-body models}

We use the \textit{Galaxia} model
\citep{sharma11} to create a synthetic survey of the Milky Way.  We
choose the default (analytical) \textit{Galaxia} model for the disc
population (the Besan\c{c}on model, \citealt{robin03}), and the
\cite{bullock05} (BJ05) N-body models for the stellar
halo. \textit{Galaxia} employs a scheme to sample the N-body models,
which ensures that the phase-space density of the generated stars is
consistent with that of the N-body particles. There are eleven stellar
halo models, each representing a different assembly history and
stellar halo mass. This suite of simulated stellar haloes have been
used extensively in the literature \citep[e.g.][]{bell08, xue11,
  deason13}, and although there may be limitations relative to more
sophisticated cosmological simulations, they are an incredibly useful
tool for testing and calibrating observational survey data.

The BJ05 models only include halo stars from accreted dwarf galaxies,
there are no halo stars born ``in-situ" in the parent halo, as
predicted by cosmological hydrodynamic simulations
\citep[e.g.][]{zolotov09,font11}. However, if this population does
exist (this is still not clear in the Milky Way: \citealt{deason17,
  belokurov18, dimatteo18,haywood18}) it is likely confined to the
inner halo and will have similar properties to the thick disc
\citep{zolotov09, mccarthy12, pillepich14, belokurov19, Gallart2019}. Thus, in our
decomposition of halo/disc populations (see Section \ref{sec:decomp})
any \textit{in-situ} halo stars will likely be labeled as disc
stars. However, we cannot exclude the possibility that some fraction
of the stellar halo mass we compute in the \textit{Gaia} data has an
\textit{in-situ} origin. This is discussed further in Section
\ref{sec:disc}.

A synthetic survey is produced from the models in Johnson-Cousins bandpasses and converted to the \textit{Gaia} photometry using the relations given in \cite{jordi10}. Uncertainties in photometry and astrometry applicable to GDR2 are also included in the model. This is implemented using the Python \textsc{PyGaia} package\footnote{\url{https://pypi.org/project/PyGaia/}}. This module implements the performance models for \gaia\, which are publicly available\footnote{\url{http://www.cosmos.esa.int/web/gaia/science-performance}}.

In the top panels (a-d) of Fig. \ref{fig:cmd} we show colour-magnitude diagram (CMDs) for high latitude ($|b| > 30^\circ$) stars in the \textit{Galaxia} model. Panels (a) and (b) show apparent magnitude vs. colour, and panels (c) and (d) show absolute magnitude vs. colour (with apparent magnitude restricted to $14 < G < 17$). The dashed lines
indicate the colour range we consider for candidate RGB stars. In panels (b) and (d), we exclude stars with parallax $>0.2$ (approx. $D < 5$
kpc). This cut removes nearby dwarf stars, but there are still disc giants present. We indicate the disc and halo populations with
the red and blue contours, respectively. We have checked that the
completeness of the halo star sample is not significantly affected by
the parallax cut. We find that, for the magnitude and colour range
under consideration, the halo stars with $D > 5$ kpc are complete to
$\geq 90\%$. Our selection of RGB stars, based on magnitude, colour
and parallax, spans the distance range $ D \sim 5-100$ kpc.  In panel (h) of Fig. \ref{fig:cmd} we show the proper motions of
the RGB stars in Galactic coordinates ($\mu_\ell, \mu_b$). Here, we only consider stars with parallax $<0.2$, $1.0 < G_{\rm BP} - G_{\rm RP} < 1.6$ and $14 < G < 17$. The disc
and halo components are indicated with the red and blue contours,
respectively. The disc and halo components have distinct, but
overlapping, proper motion distributions. Here, we are showing all
stars across the sky, but these sequences vary depending on the
Galactic coordinates (see Fig. \ref{fig:pm_disk_halo}). This figure
shows that the proper motion distributions of RGB stars can be used to
disentangle the disc and halo populations. In Section \ref{sec:decomp}
we use the proper motion distributions to estimate the number of RGB
stars in the halo in bins of colour, magnitude and position on the
sky.

In panels (e-g) of Fig. \ref{fig:cmd} we show the equivalent CMDs (using only apparent magnitude) and proper motions of the GDR2 data (see below).

\begin{figure}
  \includegraphics[width=8.5cm, angle=0]{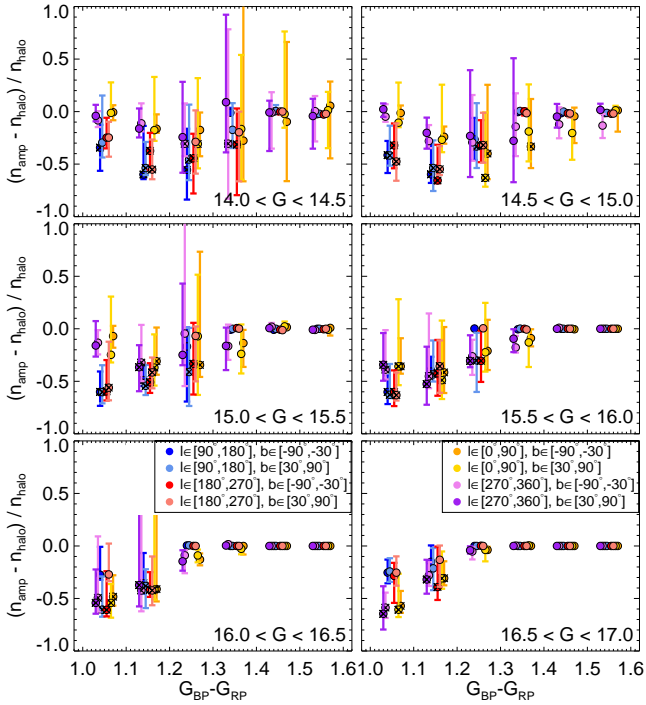}
\caption{The estimated number of halo RGB stars in the  \textit{Galaxia} models from the XD fitting ($n_{\rm amp}$) relative to the true number ($n_{\rm true}$) as a function of colour. Here, we have combined results from all eleven BJ05 haloes and show the median and 16/84 percentiles. Each panel indicates a different magnitude bin. The coloured filled circles indicate the (8) bins in Galactic longitude and latitude. The colour scheme is given in the legend (also shown in Fig. \ref{fig:area}). Bins where the amplitude is underestimated or overestimated by more than 30 percent, are shown with black crosses. In these cases, the disc and halo are difficult to distinguish, and we can exclude these bins in our analysis. However, in most of the bins (70 percent) we are able to recover the true number of halo stars to within 30 percent.}
\label{fig:namp}
\end{figure}

\subsection{Gaia DR2}
The models described in the previous sub-section are tailored towards
the GDR2 dataset. Before going further, we briefly describe our
selection of the real \gaia\ data. We select stars from GDR2 with
photometry, parallax, and proper motion information. The photometry is
corrected for extinction using the \cite{schlegel98} dust maps, and we
use the relations given in \cite{gaia_phot} to correct the $G$,
$G_{\rm RP}$ and $G_{\rm BP}$ bandpasses. We only include stars with
re-normalised unit weight error, $\mathrm{RUWE} < 1.4$ \citep{ruwe},
which ensures stars with unreliable astrometry are excluded. In
addition, we exclude stars with large BP/RP flux excess using the
selection given in \cite{gaia_phot}. These cuts remove $\sim 8\%$ of the sample in the colour, magnitude and latitude range under consideration (see below). Note most of the star excised are at the fainter, redder region of our selection. We assume that these quality cuts affect the disc and halo populations equally, and thus increase our estimated luminosity (and mass) estimate of the Milky Way (see Section \ref{sec:lumf}) by 8\%. From the cleaned sample, we
select RGB stars at high latitude ($|b| > 30^\circ$) with parallax
$<0.2$, $1.0 < G_{\rm BP} - G_{\rm RP} < 1.6$ and $14 < G < 17$ (see
Fig \ref{fig:cmd}).

In the following Section, we decompose the disc and halo RGB stars
using proper motion information. First, we illustrate this process
using the \textit{Galaxia} models, and we then apply the technique to
our GDR2 sample.

\section{Disc-Halo Decomposition}
\label{sec:decomp}
\begin{figure*}
  \begin{minipage}{0.33\linewidth}
        \centering
        \includegraphics[width=5.7cm,angle=0]{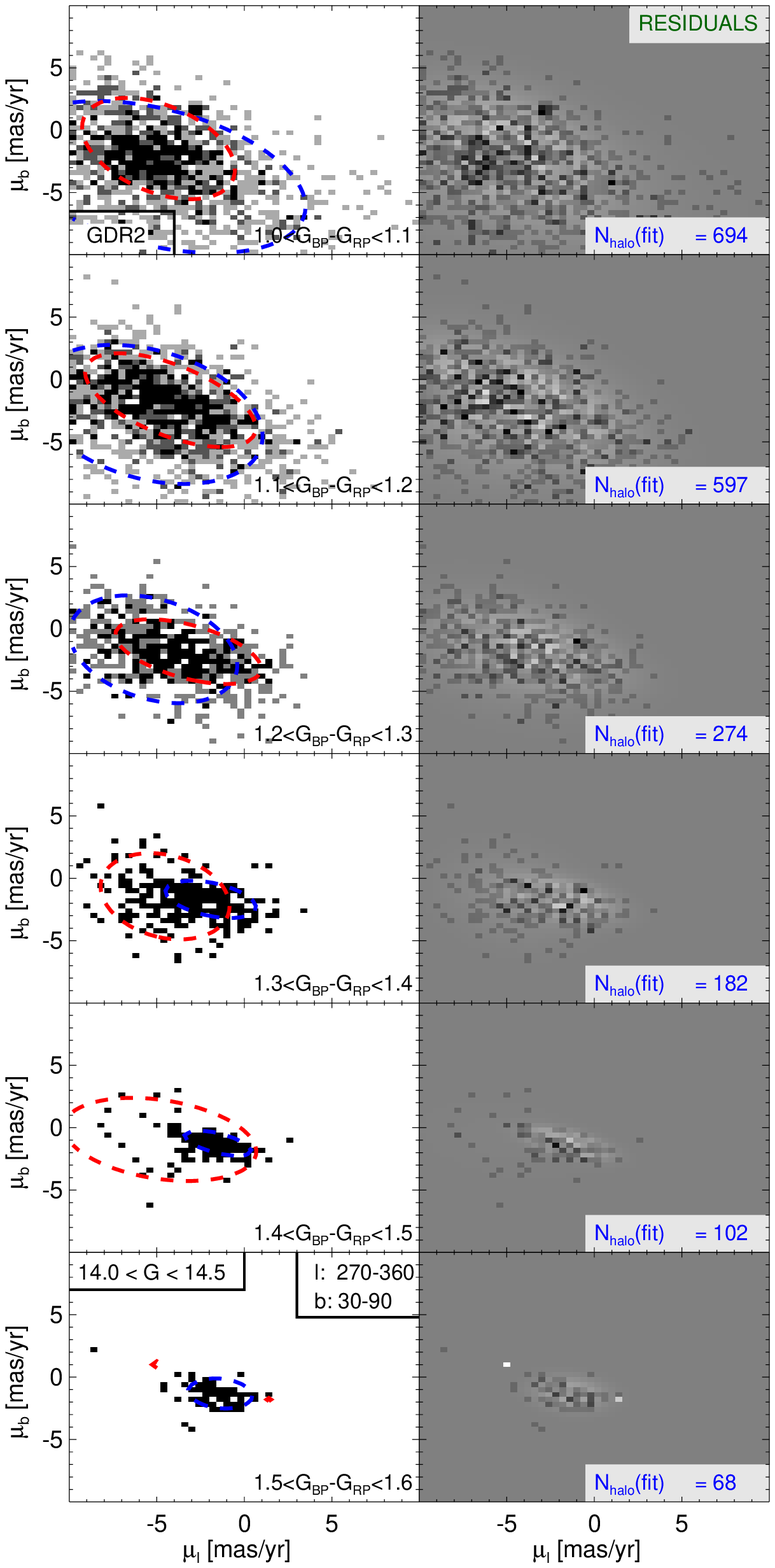}
  \end{minipage}
  \begin{minipage}{0.33\linewidth}
        \centering
        \includegraphics[width=5.7cm,angle=0]{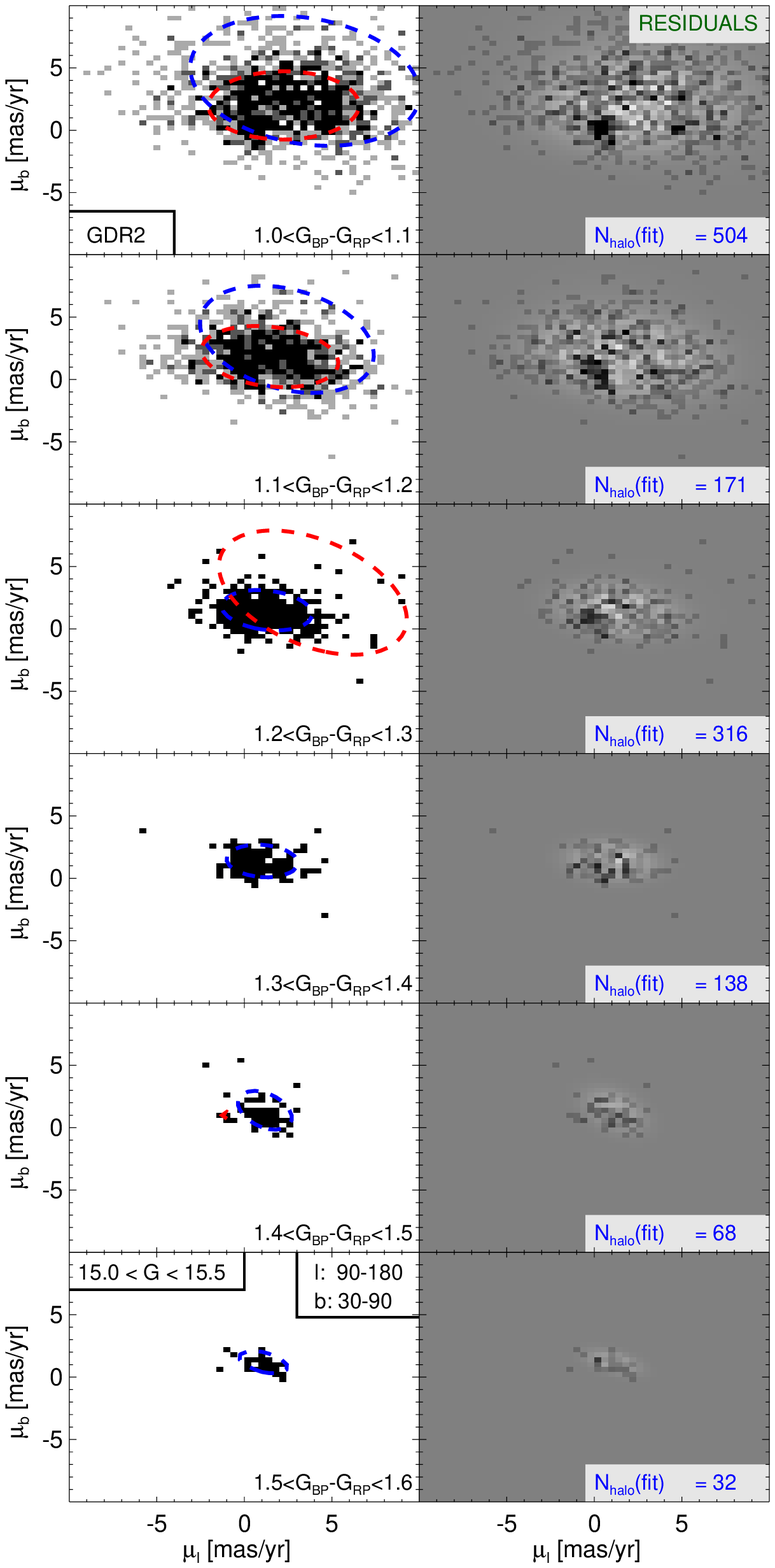}
  \end{minipage}
  \begin{minipage}{0.33\linewidth}
        \centering
        \includegraphics[width=5.7cm,angle=0]{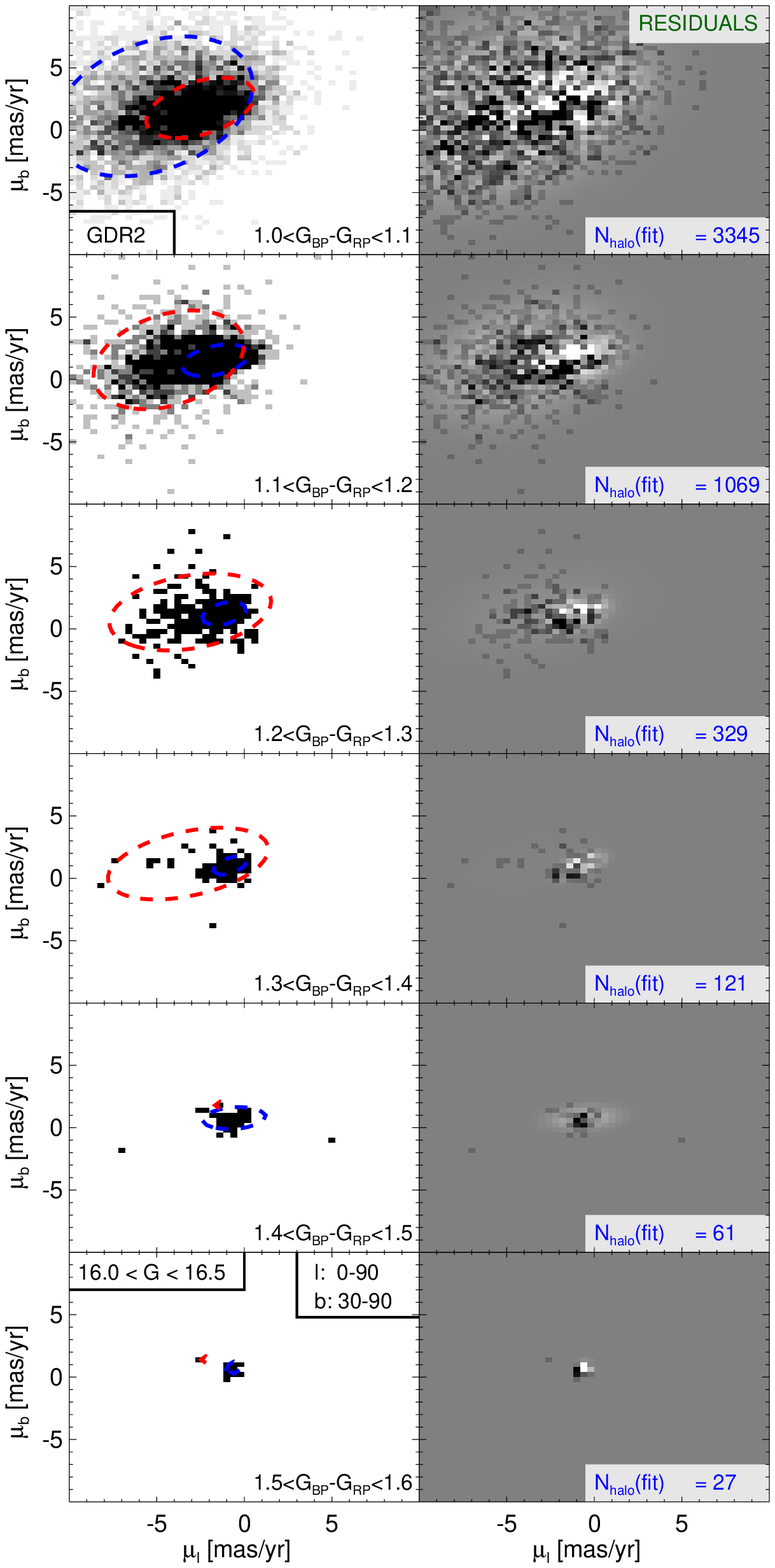}
  \end{minipage}
        \caption{\textit{Left columns:} Extreme deconvolution (XD) fits to the GDR2 proper motion distributions in bins of $G_{\rm BP} - G_{\rm RP}$ colour.  The dashed red and blue lines show the estimated disc and halo distributions. Here, we use the Galaxia model values to initialize the XD fit. \textit{Right columns:} The residuals from the fit. Here, the pixel size is 0.4 mas/yr, and the shading saturates at an excess of $\Delta = \pm 5$. The estimated number of halo stars is given in the bottom right hand corner. Three examples are shown for different magnitude ranges and bins on the sky.}
          \label{fig:resid_gdr2}
\end{figure*}

In Fig. \ref{fig:cmd} we showed that our selection of RGB stars
includes both halo and disc populations. In order, to disentangle
these populations, we use the 2-dimensional proper motion
distributions. We assume 2-D (for each component of proper motion)
Gaussian distributions for both the halo and disc. This Gaussian
approximation is reasonable as we (independently) fit in bins of
magnitude, colour and position on the sky, rather than fit the entire
distribution with one 2-D Gaussian.  We use 6 bins in magnitude
(between $14 < G < 17$), 6 bins in colour (between $1.0 < G_{\rm BP} -
G_{\rm RP} < 1.6$), and 8 spatial bins. The spatial bins are shown in
Fig. \ref{fig:area}. When applying this method to the \gaia\ data we
exclude stars within 30 deg of the LMC and 10 deg of the SMC. We also
perform the analysis both with and without stars in the vicinity of
the Sagittarius (Sgr) stream. The Sgr stars are selected to lie within 12 degrees of the tracks shown in Fig. \ref{fig:area} \citep[see][]{deason12, belokurov14}. Note that when we use the BJ05 stellar halo models we do not attempt to excise any streams or satellites, so the biases from unrelaxed substructures are likely more pronounced in the models than the data.

In Fig. \ref{fig:pm_disk_halo} we show the true Gaussian parameters for the disc and halo populations in the \textit{Galaxia} model. For this illustration the halo component is Halo-7, although similar trends are seen in all of the haloes. This figure shows that the overlap between the disc and halo components varies as a function of magnitude, colour and position on the sky. In some cases the overlap is larger, and in others the populations are more distinct. To perform the fits simultaneously (i.e. without knowing which stars belong to disc or halo), we model the proper motion distributions with a mixture of two (halo+disc) multi-variate Gaussians using the Extreme Deconvolution algorithm described in \cite{bovy11}. In Fig. \ref{fig:resid} we show the outcome of these fits for the \textit{Galaxia} model. Note that we initialise the fits using the true Gaussian values for the disc and a halo model (Halo-7 in this case). This step is taken to avoid misclassification of the halo/disc components. However, we check that initialising with different halo profiles, or an independent disc model makes little difference to the results (see later). Fig. \ref{fig:resid} shows that in some bins the decomposition works well, while in others we are unable to clearly distinguish the distinct components. We note the true and fitted halo amplitudes (number of halo stars) in the bottom right corner of the panels. Bins at redder colours and fainter magnitudes have little, if any, disc component so the fits are straightforward. However, even with a significant disc contribution (e.g. at bluer colours, and brighter magnitudes) we can sometimes get good estimates of the halo amplitudes.

The reliability of the decomposition for each bin is illustrated in Fig. \ref{fig:namp}. Here, we show the relative difference between the estimated and true number of halo stars. We have combined results from all eleven BJ05 haloes and show the median and 16/84 percentiles. In certain bins, our estimates are poor (over/under estimate by more than 30 percent) and these are shown with the black crosses. These are cases where the overlap between disc and halo makes decomposition based on proper motion alone very difficult.  However, in most of the bins (70 percent) we are able to recover the true number of halo stars to within 30 percent. When we apply this method to the \gaia\ data we can exclude the bins with significant systematics. 

\begin{figure*}
        \centering
        \includegraphics[width=17cm,angle=0]{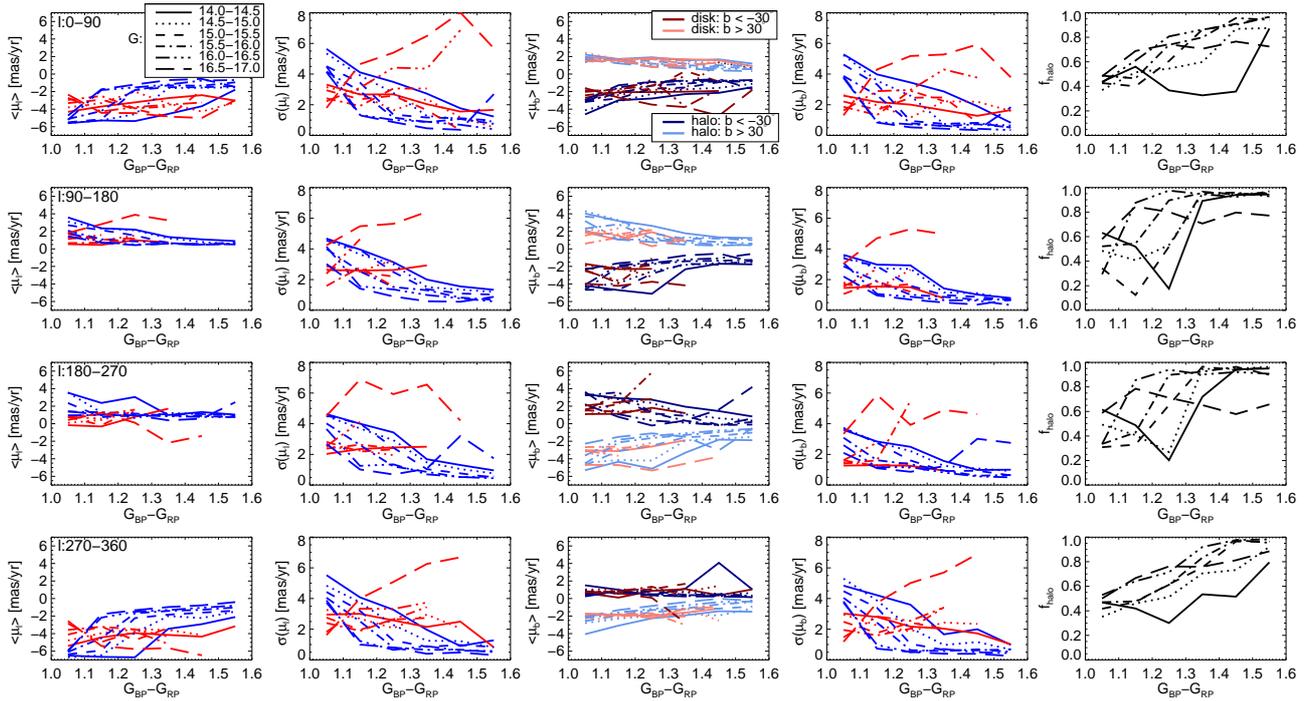}
        \caption{The mean (first and third panels) and dispersion (second and fourth panels) of the GDR2 model proper motions in Galactic coordinates as a function of $G_{\rm BP} - G_{\rm RP}$ colour. Blue and red lines indicates the estimated halo and disc components, respectively. Different magnitude bins are shown with different linestyles, and each row shows a different bin in Galactic longitude. The sequences are very similar for bins above and below the Galactic plane, except for $\langle \mu_b \rangle$ (third column), which we indicate with different shades of blue and red. The last panel on the right indicates the fraction of halo stars as a function of colour. The sequences follow roughly the expected trends (see Fig. \ref{fig:pm_disk_halo}). However, the ``disc" component in the faintest magnitude bin appears to be dominated by noise. }
          \label{fig:pm_disk_halo_gdr2}
\end{figure*}

In Fig. \ref{fig:resid_gdr2} we show examples of the 2D Gaussian fits to the \gaia\ data. These example bins are the same as in Fig. \ref{fig:resid}. We show the more general results in Fig. \ref{fig:pm_disk_halo_gdr2}. Here, we can see the resulting Gaussian parameters behave similarly to the model predictions (shown in Fig. \ref{fig:pm_disk_halo}).  We note that a noise component becomes apparent in the faintest bins ($16.0 <G < 17$), which is labeled as ``disc''. This component is relatively minor, as the number of stars belonging to the disc in the faint, red bins is very low ($N \lesssim 50$). Moreover, in all bins, the halo component appears to be well-behaved, which gives us confidence that our estimated halo amplitudes are reasonable. 

Figures \ref{fig:resid_gdr2} and \ref{fig:pm_disk_halo_gdr2} also help us evaluate one of the assumptions we have made in our modeling: that the proper motion is a reliable distance indicator. In essence, we are using proper motion to disentangle distant halo stars and nearby disc stars. However, populations such as the thick disc or \textit{in-situ} halo could potentially break this decomposition if their proper motion distributions mimic the (accreted) halo. The results of the decomposition give us confidence that this is not the case. First, the general trends seen in Fig. \ref{fig:pm_disk_halo_gdr2} look similar to the model predictions shown in Fig. \ref{fig:pm_disk_halo}. Note the agreement is even better when we compare with the ``fitted'' values for the model, rather than the ``true'' values. This agreement is non-trivial: it shows that the inferred halo population in GDR2 resembles the accreted halo population in the models. If thick disc or \textit{in-situ} halo stars were contaminating the results, the proper motions distributions would be inflated (because these stars are closer), and wouldn't necessarily narrow with colour and magnitude, as seen in Fig. \ref{fig:pm_disk_halo_gdr2}. Second, the redder bins (see e.g. lower panels of Fig. \ref{fig:resid_gdr2}) appear to be almost entirely comprised of very distant stars with small proper motions. If a significant fraction of thick disc or \textit{in-situ} halo stars were contaminating these bins the proper motion distributions would be much broader. However, we caution that we can not exclude the possibility that our halo sample includes any in-situ material, particularly if these stars can reach out to large distances. This is discussed further in Section \ref{sec:disc}.

To provide error estimates on the number of halo RGB stars in each bin, we perform the fits $N=100$ times. Before each fit, we scatter the parallax according to the error distribution and then make a cut of parallax $< 0.2$. This step essentially adds/removes stars from the analysis with parallax close to the limiting threshold. In addition, we randomly select one of the eleven BJ05 haloes to initialise the fits. As a final check, we initialise the disc parameters using a completely independent model to \textit{Galaxia}. For this we use the disc model described in \cite{sanders15}. This model has an action distribution that varies smoothly with age and metallicity using analytic prescriptions for dynamical heating, radial migration and the radial enrichment of the interstellar medium over time. A mock catalogue of on-sky position, magnitude, age, metallicity, mass and velocities was generated using Markov Chain Monte Carlo sampling \citep{foreman13} of the model combined with a set of PARSEC isochrones. We require samples to have $14<G<17$ and $1<(G_\mathrm{BP}-G_\mathrm{RP})<1.6$, and convolve the output samples in parallax, proper motion and magnitudes using nearest neighbours in magnitude and on-sky position from GDR2. We find that, after initialising the disc component using the \cite{sanders15} model, the resulting halo amplitudes are very similar, and do not significantly affect our derived luminosity (see following section).
 
\section{Total stellar halo Luminosity}
\label{sec:lumf}
\begin{figure}
  \includegraphics[width=8.0cm, angle=0]{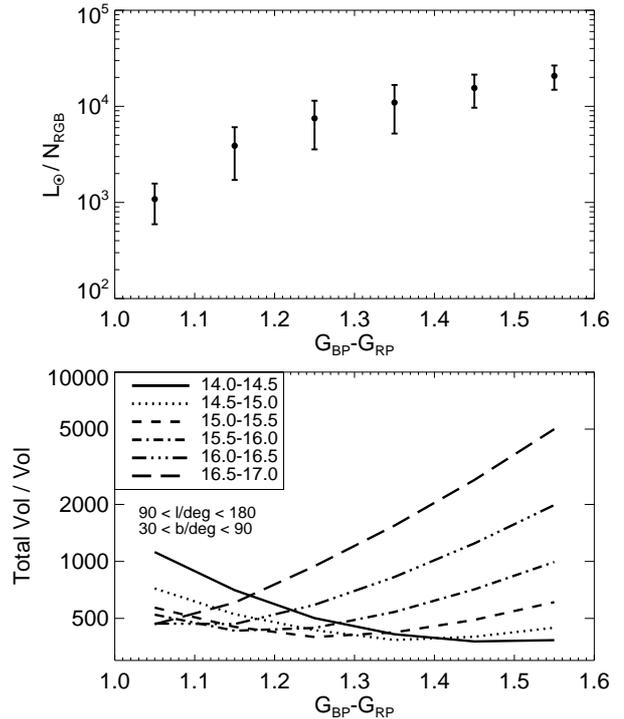}
\caption{\textit{Top:} The relation between total luminosity and number of RGB stars per colour bin. Here, we have used a set of weighted PARSEC isochrones assuming uniform ages in the range 10-14 Gyr, and a metallicity distribution with $\langle \mathrm{[Fe/H]} \rangle = -1.5$, $\sigma(\mathrm{[Fe/H]})=0.5$. \textit{Bottom:} The total volume (out to 100 kpc) relative to the volume probed by a colour bin. Different linestyles correspond to different magnitude bins. Here, we used the weighted isochrones to estimate the distance range probed by a specific colour, magnitude bin, and we use the stellar halo density profile to relate the volume probed to the total volume. For the Milky Way, we assume an Einasto profile with $n=1.7$, $R_e =20$ kpc and a minor-to-major axis ratio $q=0.6$ \citep{deason11}.}
\label{fig:lum_vol}
\end{figure}

In the previous Section, we calculated the number of halo RGB stars in bins of colour, magnitude and regions on the sky. We now want to convert these numbers into an estimate of the total stellar halo luminosity (and hence stellar mass). We provide a luminosity estimate for each bin, by applying the following two corrections:

\begin{enumerate}

\item \textit{Stellar population correction:} We use isochrones to relate the number of RGB stars in a given colour bin to the total luminosity. Here, we use the PARSEC isochrones \citep{bressan12}, with metallicities in the range $-2.5 <$ [M/H] $<0.0$  and ages 10-14 Gyr. These isochrones are solar scaled, but halo stars are alpha enhanced with $[\alpha/\mathrm{Fe}] \sim 0.3$ \citep[e.g.][]{venn04}. Hence, we use the relation given by \cite{salaris05} to relate [M/H] to [Fe/H]: $\mathrm{[M/H]} = \mathrm{[Fe/H]}+0.2$ for $[\alpha/\mathrm{Fe}]=0.3$.  For each isochrone, we calculate the number of RGB stars per unit luminosity as a function of colour. We adopt the PARSEC isochrones as our ``fiducial" stellar population model (these are also the models used in \textit{Galaxia}), and we comment on the changes to our results when other models are used in Section \ref{sec:disc}.  For each of our 6 colour bins (with 0.1 dex width) we calculate $N_{\rm RGB}/L_\odot$. This procedure requires us to assume an initial mass function (IMF):
\begin{equation}
\frac{N_{\mathrm{RGB}, i}}{L_\odot} = \frac{\int^{m_2}_{m_1} \xi(m) \, \mathrm{d} m}{\int^{m_{\rm max}}_{m_{\rm min}} L \, \xi(m) \, \mathrm{d} m}
\end{equation}
where, $\xi(m)$ is the IMF and $i$ denotes the isochrone. The limits
$m_1$ and $m_2$ give the mass range probed by a particular colour bin,
and $m_{\rm min}$, $m_{\rm max}$ denotes the full range of masses
probed by the isochrone. Note that the luminosity estimate is only
weakly dependent on the IMF, as most of the commonly used IMF
parameterizations are very similar for the high mass stars, which
dominate the stellar light. In comparison, the stellar mass strongly
depends on the adopted IMF, as the uncertainty of the mass function
for low mass stars, which dominate the mass, is significant. It is for
this reason that we provide a robust estimate of total stellar
luminosity, rather than mass. This luminosity can later be converted
to stellar mass using the appropriate stellar mass-to-light ratio for
a given IMF (see Section \ref{sec:disc}). For the \textit{Galaxia}
models we use a Chabrier IMF (\citealt{chabrier03}, as assumed for the
halo's N-body component in this model), and we use the Kroupa IMF
\citep{kroupa01} when estimating the Milky Way halo luminosity using
\gaia\ data. In practice, these IMFs are comparable and give very
similar luminosity (and mass) estimates.

We next convert $N_{\mathrm{RGB},i}/L_\odot$ for each isochrone
into an overall estimate by weighting the isochrones using a
metallicity distribution function and age distribution. For the \textit{Galaxia} models we fit a Gaussian to the true MDF of
the halo, and for the \gaia\ data we adopt an MDF from the literature with $\langle \mathrm{[Fe/H]} \rangle = -1.5$,
$\sigma(\mathrm{[Fe/H]})=0.5$ \citep{an13,zuo17}. For the ages, we
assume a uniform age distribution in the range 10-14 Gyr. The top
panel of Fig. \ref{fig:lum_vol} shows the resulting (weighted)
$N_{\mathrm{RGB}}/L_\odot$ for each colour bin. The error bars
indicate the 16/84 percentiles given the adopted MDF and age
distribution. We now have a way to relate total number of RGB stars in
a colour bin to the luminosity. However, our estimates from the
previous section are in bins of magnitude and area on the sky, and
thus each probe a different volume of the halo. Thus, the final
correction is to volume correct each bin.

\item \textit{Volume correction:} Each bin in magnitude, colour and region of the sky probes a different volume of the halo. Thus, to convert our estimated number of halo RGB stars to total number of RGB stars we need to volume correct. This requires adopting a density profile for the stellar halo. This has been measured for the Milky Way in previous work, and we adopt an Einasto profile when applying to the \gaia\ data, with $n=1.7$, $R_e = 20$ kpc and 
minor-to-major axis ration $q=0.6$ \citep{deason11}. For the \textit{Galaxia} models we fit an Einasto profile directly to the halo stars out to 100 kpc. For all eleven haloes, the values typically lie in the range: $n=1-5$, $R_e =15-40$ kpc and $q=0.4-0.8$. Our volume correction relates the volume probed by each bin to the total volume, which we assume goes out to 100 kpc. Hence, our luminosity estimates are within 100 kpc, although this is more or less identical to the total luminosity as there is very little stellar halo mass beyond 100 kpc. We use the PARSEC isochrones to calculate the distance range probed in each bin, and by adopting a density profile this can be converted into a volume:
\begin{equation}
\frac{\mathrm{Vol,i}}{\mathrm{Total \, Vol}} = \frac{\int^{D_2}_{D_1} \int^{\ell_2}_{\ell_1} \int^{b_2}_{b_1} \rho (D, \ell, b) \, D^2 \, \mathrm{cos}(b)\, \mathrm{d} D \mathrm{d} \ell \mathrm{d} b}{\int^{D=100 \rm kpc}_{D=0 \rm kpc} \int^{\ell=360^\circ}_{\ell=0^\circ} \int^{b=90^\circ}_{b=-90^\circ} \rho (D, \ell, b) \, D^2 \, \mathrm{cos}(b) \, \mathrm{d} D \mathrm{d} \ell \mathrm{d} b}
\end{equation}
where, $i$ denotes an individual isochrone and $D_1, D_2, \ell_1, \ell_2, b_1, b_2$ denote the range in distance and area probed by each bin (where the minimum value of $D_1 = 5$ kpc). The combined estimates are then calculated by weighting  the isochrones by  an MDF and age distribution.  In the bottom panel of Fig. \ref{fig:lum_vol} we show this volume correction for one bin in $\ell$ and $b$ as a function of magnitude and colour.
\end{enumerate}

\begin{figure*}
  \begin{minipage}{\linewidth}
        \centering
        \includegraphics[width=18cm,angle=0]{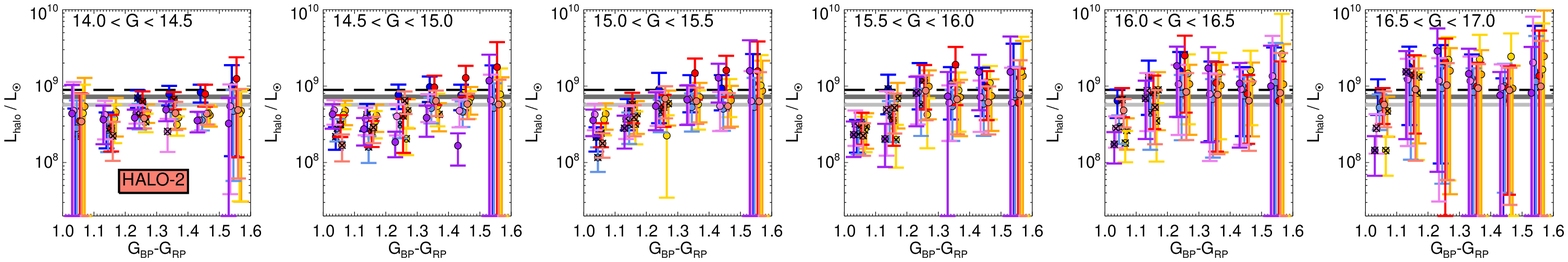}
  \end{minipage}
  \begin{minipage}{\linewidth}
        \centering
        \includegraphics[width=18cm,angle=0]{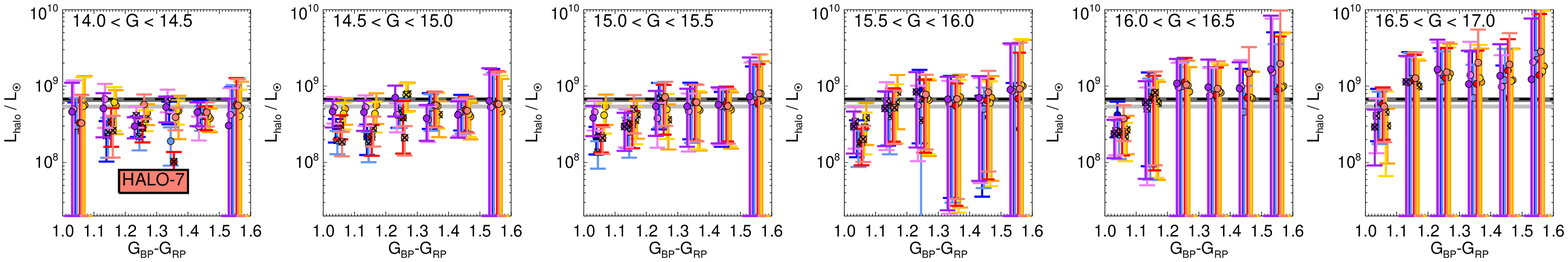}
  \end{minipage}
  \begin{minipage}{\linewidth}
        \centering
        \includegraphics[width=18cm,angle=0]{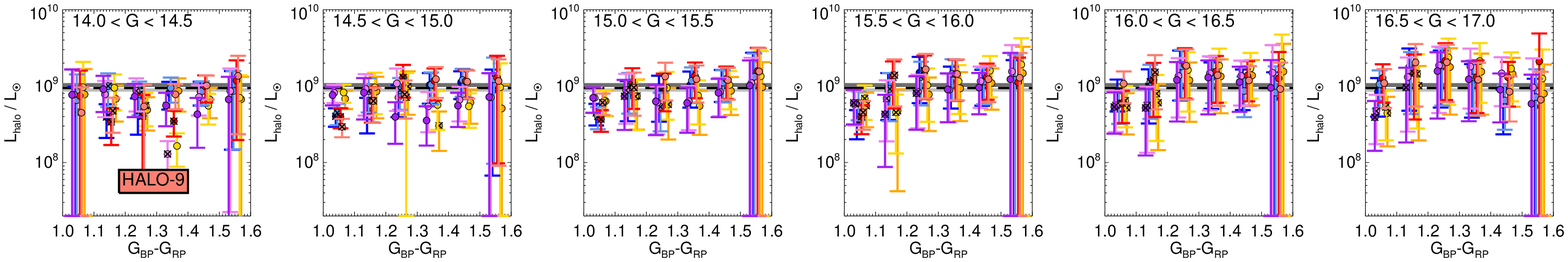}
  \end{minipage}
        \caption{The estimated (total) stellar halo luminosity as a function of colour. Each panel shows a different magnitude bin. For each colour, magnitude bin, there are 8 bins on the sky. The colour coding is the same as in Fig. \ref{fig:area}. We show three example haloes from the Galaxia+N-body models. The dashed black line shows the true value, and the (light) shaded grey region indicates the combined estimate from all of the colour, magnitude and $(\ell, b)$ bins. The dark shaded grey region indicates the combined estimated when 30 percent of the bins (shown with black crosses) are excluded.}
          \label{fig:lum_bins}
\end{figure*}

After applying the corrections outlined above we can estimate the total stellar halo luminosity. First, we test the method on the \textit{Galaxia} models, for which we know the true halo luminosity. In Fig. \ref{fig:lum_bins} we show the estimated luminosity for every bin in colour (x-axis), magnitude (panel) and area on the sky (coloured symbols) for three example haloes. The light grey region indicates the combined estimate for all bins, and the dark grey region indicates the combined estimate for selected bins. These selected bins are identified in the previous section, and exclude bins where the overlap between disc and halo prevents a good estimate of the number of halo RGB stars. Here, approximately 30\% of the bins are excluded and these are indicated with the black crosses in the figure. The black dashed line indicates the true halo luminosity (out to 100 kpc). The luminosity estimates in each bin have large error bars, but the combination of a large number of these bins can give a $\sim 5\%$ measure (but note this error is just statistical!). Reassuringly, the estimates in different bins generally agree very well, apart from the bins that we have already identified as having systematic differences (black crosses).

\begin{figure}
  \includegraphics[width=8.5cm, angle=0]{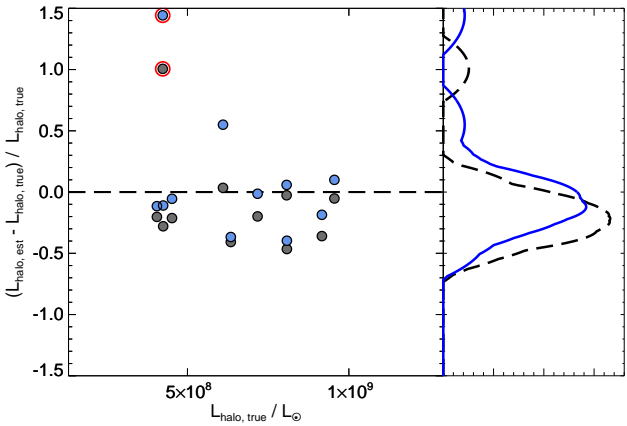}
\caption{The estimated luminosity for the Galaxia+N-body haloes relative to the true values as a function of stellar halo luminosity. Here, the ``total'' luminosity is defined within 100 kpc. The right-inset panel shows the PDF for the $\left(L_{\rm halo, est} - L_{\rm halo, true} \right)/L_{\rm halo,true}$ values. The grey points are the estimates when all bins are used, and the blue points are when bins with high levels of contamination are excluded. For the majority of haloes we can recover the true value to within $\sim 25\%$. An outlier (halo-10) is indicated with a red circle; this halo has significant contribution from unrelaxed substructure.}
\label{fig:lum_pdf}
\end{figure}

The results for all eleven of the BJ05 haloes are shown in Fig. \ref{fig:lum_pdf}. Here, we show the estimated luminosity relative to the true luminosity. The grey filled circles show the combined estimates from all bins, and the blue filled circles show the combined estimates from a subset of ``robust'' bins. The luminosity is typically underestimated by 20\% when all bins are used. This is because for certain bins the halo and disc populations cannot be properly decomposed. However, if we disregard these bins we are able to recover the true value to within 25\%. Note the scatter across all eleven haloes is larger than the individual statistical error bars ($\sim 5\%$). This is due to systematic effects, such as substructure, non-gaussian MDFs, non-Einasto density profiles etc. So, this exercise gives us a more robust estimate of the error of our estimated luminosity.

\begin{figure*}
  \begin{minipage}{\linewidth}
        \centering
        \includegraphics[width=18cm,angle=0]{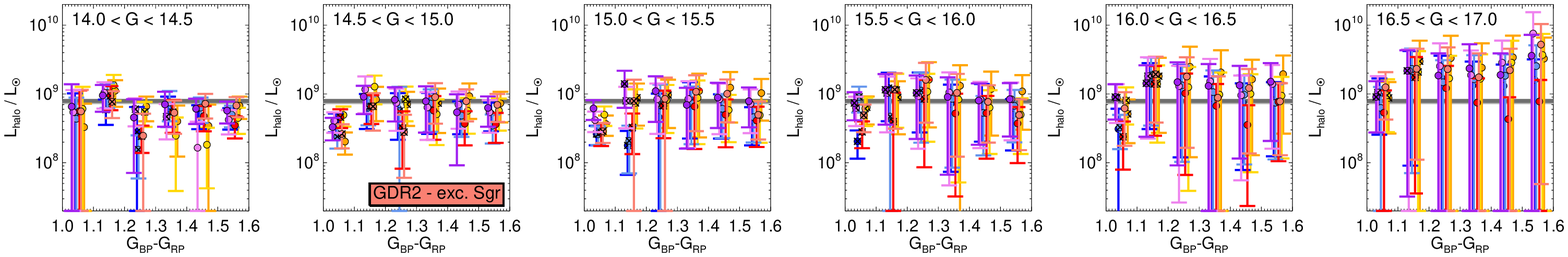}
  \end{minipage}
  \begin{minipage}{\linewidth}
        \centering
        \includegraphics[width=18cm,angle=0]{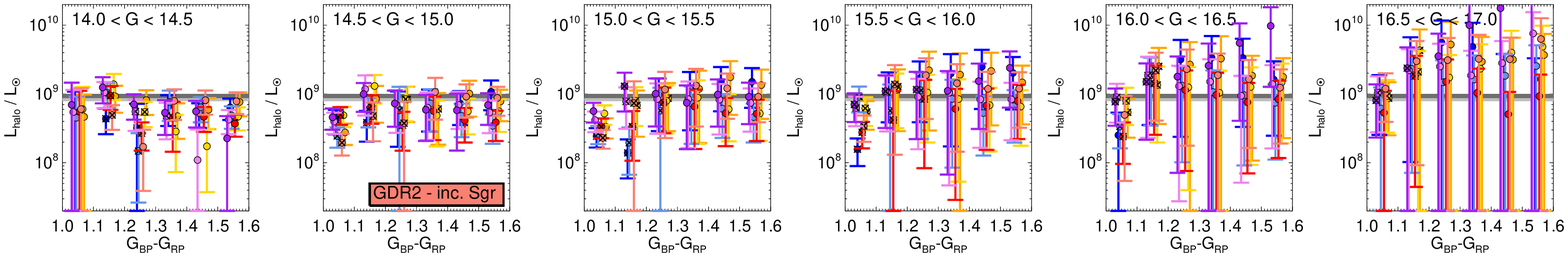}
  \end{minipage}
        \caption{The estimated (total) stellar halo luminosity as a function of colour. Each panel shows a different magnitude bin. For each colour, magnitude bin, there are 8 bins on the sky. The colour coding is the same as in Fig. \ref{fig:area}. The top and bottom rows show cases with Sgr excluded (top, $L_{\rm halo} = 7.9 \times 10^8 L_\odot$) and included (bottom, $L_{\rm halo} = 9.4 \times 10^8 L_\odot$). Including Sgr increases the total luminosity by $\sim 15\%$. The (light) shaded grey region indicates the combined estimate from all of the colour, magnitude and $(\ell, b)$ bins. The dark shaded grey region indicates the combined estimated when 30 percent of the bins (shown with black crosses) are excluded.}
          \label{fig:gdr2_lum_bins}
\end{figure*}

We now apply our procedure to the \gaia\ data and show the results for the luminosity estimate in Fig. \ref{fig:gdr2_lum_bins}. Here, we have performed the analysis both with and without the Sgr stream. When the Sgr stream is included the estimated luminosity increases by 15\%. It is clear that including Sgr enhances the halo luminosity, particularly in the fainter, redder bins. This is particularly evident in the $\ell \in [270,360], b \in [30,90]$ bin, which is where the apocentre of the Sgr leading arm (at $D \sim 50$ kpc) is dominant. These results give a rough estimate of the Sgr luminosity of $L_{\rm Sgr } \sim 1.5 \times 10^8 L_\odot$, in good agreement with the value derived by \cite{ostholt12}. Owing to the systematics we deduced in the previous section we use a subset of bins to calculate our best luminosity estimate. We find $L_{\rm halo}=7.9 \pm 2.0 \times 10^8 L_\odot$ excluding Sgr, and  $L_{\rm halo}=9.4 \pm 2.4 \times 10^8 L_\odot$ including Sgr. Here, we have assumed, based on comparison to N-body models, that this estimate is accurate to 25\%. Note that if we had used all available bins, our estimates are reduced by $\sim 10\%$, and the statistical error is smaller. However, as shown in Fig. \ref{fig:lum_pdf}, the systematic error increases, and the mass is likely underestimated when all bins are used.

\section{Discussion}
\label{sec:disc}

\subsection{A relatively high Galactic stellar halo mass?}
In the preceding Section(s) we have used counts of RGB stars in GDR2 to estimate the total luminosity of the Galactic halo. This can be converted to a stellar mass by adopting an appropriate stellar mass-to-light ratio. Using the (weighted) suite of PARSEC iscohrones described earlier (with uniform ages between $10-14$ Gyr, and an MDF with $\langle \mathrm{[Fe/H]} \rangle = -1.5$), we estimate stellar mass-to-light ratios of 1.3, 1.5 and 2.8 for a Chabrier, Kroupa and Salpeter \citep{salpeter55} IMF, respectively. We adopt the Kroupa IMF as our fiducial model, which gives: $M^\star_{\rm halo} = 1.2 \pm 0.3 \times 10^{9}M_\odot$ (exc. Sgr) and $M^\star_{\rm halo} = 1.4 \pm 0.4 \times 10^{9}M_\odot$ (inc. Sgr). Alternatively, we can express these values in terms of the local stellar halo density: $\rho_0 = 6.9 \times 10^{-5} M_\odot/\mathrm{pc}^3$ (exc. Sgr), $\rho_0 = 8.1 \times 10^{-5} M_\odot/\mathrm{pc}^3$ (inc. Sgr). These values can be multiplied by factors of $1.3/1.5$ or $2.8/1.5$ if Chabrier or Salpeter IMFs are preferred.

Our estimated stellar mass is significantly higher than recent values in the literature. For example, \cite{bell08} and \cite{deason11} find masses $M^\star_{\rm halo} =3-4 \times 10^8M_\odot$, which, even with the additional few $\times 10^8 M_\odot$ of substructures that are likely not accounted for in these models, is a factor of 2-3 lower than our estimate. However, it is worth pointing out that both of these estimates rely on an approximate relation between number of blue horizontal branch or main sequence turn-off stars and luminosity. These works use globular clusters to calibrate this relation, but there is no simple way to quantify the sources of systematic errors in this approach. Indeed, although the low mass quoted by \cite{bell08} and \cite{deason11} are often cited in the literature, the estimates are relatively ``back of the envelope'', and were not the main focus of the papers. All studies estimating the stellar halo luminosity or mass (including this one) face the difficult problem of converting number counts of (tracer) stars to a luminosity. The main advantages of our new estimate are (1) the uninterrupted all-sky, large volume probed by \gaia\, and (2) a thorough exploration of the various systematic uncertainties, including using simulations to test the method, the influence of the adopted IMF and stellar isochrones, and the influence of the adopted stellar density profile and MDF (see following subsection). It is intriguing that our estimate is in better agreement with results deriving from relatively nearby halo star counts \citep[e.g.][]{morrison93, gould98, dejong10}, but these require significant extrapolation to convert to a \textit{total} stellar halo mass.  Importantly, our estimated mass is in good agreement with the recent result posted by Mackereth et al. (in prep). These authors find $M^\star_{\rm halo} = 1.3^{+0.3}_{-0.2} \times 10^9 M_\odot$ using RGB star counts in APOGEE DR14 data.

It is worth remarking that the low (few $\times 10^8M_\odot$) stellar
halo mass often quoted for the Milky Way is also at odds with recent
results from \gaia\ suggesting the (inner) halo is dominated by an
ancient, massive accretion event with $M^\star \sim 0.5-1
\times 10^9M_\odot$ \citep{belokurov18, helmi18}. Moreover, analyses
of the kinematics and ages of the Galactic globular cluster
populations point to a small number of massive ($\sim 10^9M_\odot$)
Milky Way progenitors \citep{myeong18,kruijssen19}. While it is
feasible that some of the stars from these massive progenitors end up
in the stellar disc (and thus avoid being accounted for in analysis
given the $|b|>30^{\circ}$ cut), the majority of the debris should be
in the halo. Thus, our new estimate of $M^\star_{\rm halo} \sim
10^9M_\odot$ agrees with the emerging picture of a massive progenitor
dominating the stellar halo mass, and, importantly, provides a direct accounting of the debris from this event.

\subsection{Model assumptions and systematic uncertainties}
\begin{figure*}
 \begin{minipage}{0.48\linewidth}
        \centering
        \includegraphics[width=8cm,angle=0]{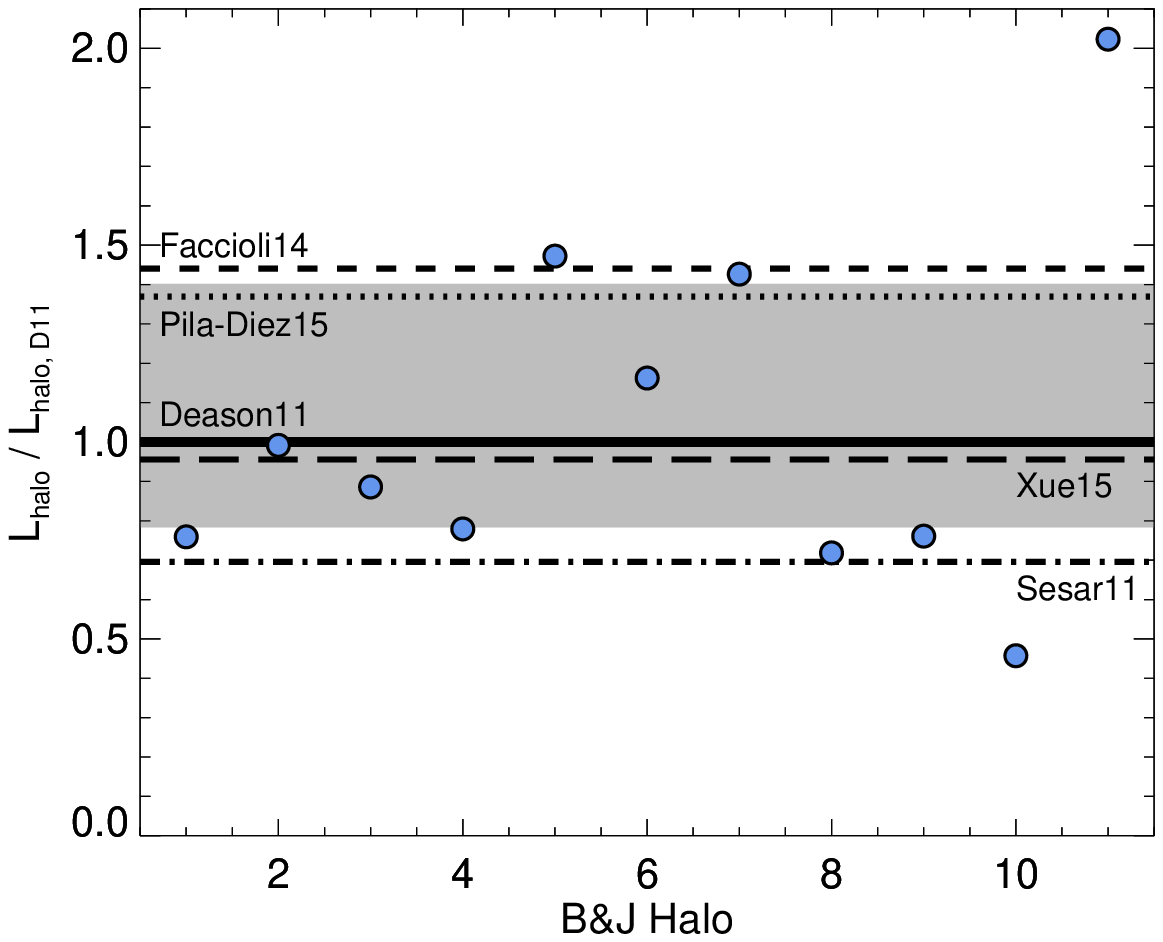}
  \end{minipage}
  \begin{minipage}{0.48\linewidth}
        \centering
        \includegraphics[width=8cm,angle=0]{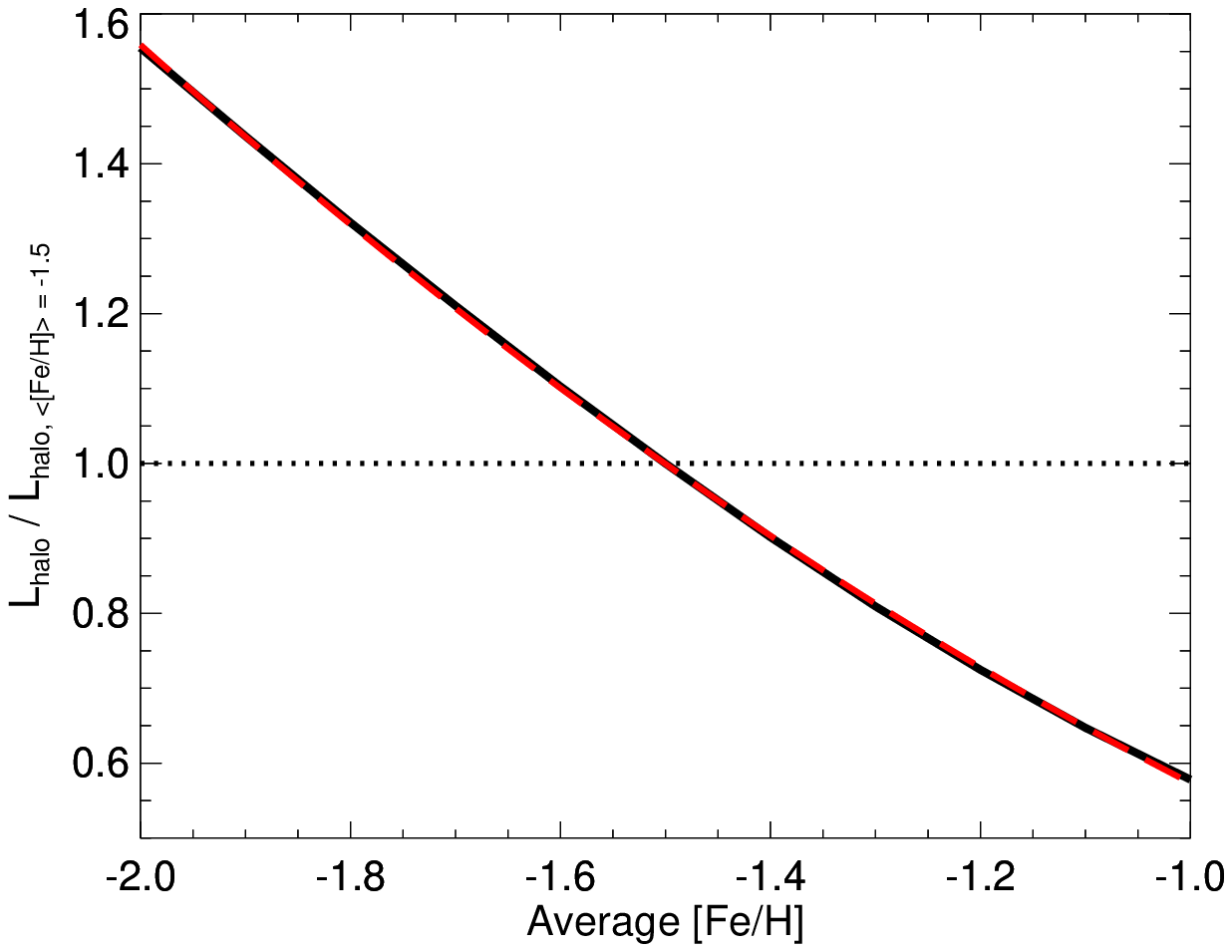}
  \end{minipage}
\caption[]{\textit{Left:} The total halo luminosity derived with various stellar halo density profiles relative to the fiducial density profile assumption. Note here we adopt a single isochrone model with age $T=10$ Gyr and metallicity [Fe/H]$=-1.5$. We use the range of density profiles seen in the BJ05 haloes (filled blue points), and indicate the results for a range of observed profiles in the literature \citep{deason11, sesar11, faccioli14, piladiez15,xue15}. Our fiducial density profile assumption \citep{deason11} lies in the middle of the estimates, but the various profiles have a 1-$\sigma$ dispersion of $\sim 30\%$ around the fiducial value. \textit{Right:} The halo luminosity relative to the fiducial luminosity (with $\langle \mathrm{[Fe/H]} \rangle =-1.5$) as a function of average metallicity. The derived luminosity strongly depends on the MDF. The dashed red line shows a quadratic fit that can be used to approximately convert our fiducial luminosity estimate to a different MDF.}
\label{fig:lum_dens_met}
\end{figure*}
In this subsection, we explore the systematic uncertainties related to our model assumptions. First, we consider the halo density profile. We adopt a flattened Einasto stellar halo density profile from \cite{deason11} to volume correct the RGB star counts in magnitude, colour and spatial bins. The form of the density profile of halo stars within 50 kpc from various sources in the literature are in broad agreement \citep[e.g.][]{faccioli14, piladiez15,xue15}, but they differ in detail. In the left-hand panel of Fig. \ref{fig:lum_dens_met} we compute the total halo luminosity for various different density profiles. Note, here for ease of computation, and as we are interested in relative differences, we adopt a single isochrone model with age $T=10$ Gyr and metallicity [Fe/H]$=-1.5$. The filled points use the density profiles relevant for the eleven BJ05 halo models. These points are shown to illustrate the range of values that can be found if there is little knowledge about the halo density profile. In this case, the dispersion in the luminosity estimates (neglecting the obvious outlier) is $\sim 35\%$ of the mean. Note we checked that the outliers in the BJ05 haloes have rather extreme density profile parameters (at least relative to the MW). The lines indicate various density profiles in the literature \citep{deason11, sesar11, faccioli14, piladiez15,xue15}. These observed profiles have been computed using a range of tracers (blue horizontal branch, RR Lyrae, main sequence, and giant stars) and data sources. This figure illustrates that the derived luminosity can vary significantly with the adopted density profile. In general, profiles that are steeper (at large distances) lead to higher luminosity estimates, as the volume correction factor  is larger. The grey region in Fig. \ref{fig:lum_dens_met} indicates the approximate $1-\sigma$ dispersion in luminosity for the range of observed density profiles, which is $\sim 30\%$ of our fiducial result using the \cite{deason11} profile.  We note that is reassuring that the profile by \cite{xue15}, which uses RGB stars as tracers, gives a similar answer to our fiducial result. We keep the \cite{deason11} profile to give our main result, but note that an additional systematic error (of 30\%) can be included in order to account for uncertainties in the stellar halo density profile.

An additional model assumption is the adopted metallicity distribution function. In our fiducial results, we adopt a MDF for the halo with $\langle \mathrm{[Fe/H]}\rangle =-1.5$. This value is motivated by results in the literature \citep{an13, zuo17}, but lower and higher average metallicities have also been reported \citep[e.g.][]{xue15, conroy19}. To explore the effect of the MDF on our results, we show in the right-hand panel of Fig. \ref{fig:lum_dens_met} the derived luminosity as a function of average metallicity. Here, we keep the same dispersion in the MDF ($\sigma =0.5$ dex) but vary the mean ($\langle \mathrm{[Fe/H]} \rangle$). Note that the \cite{deason11} density profile is adopted, but the same relative trend is seen with different stellar halo density profiles. The relation is shown relative to the fiducial luminosity assuming $\langle \mathrm{[Fe/H]} \rangle = -1.5$. The figure shows that the luminosity estimate is dependent on the adopted MDF. The adopted metallicity affects the derived total luminosity in two main ways: (1) higher metallicity isochrones have lower luminosity per unit number of RGB stars ($L_\odot/N_{\rm RGB})$, and (2) the distances, in a given colour and magnitude bin are lower at higher metallicity, and hence the volume correction (Total Vol / Vol) is typically smaller. These two effects both lead to a reduction in total luminosity at higher metallicities (and an increase at lower metallicities). To account for the variation with metallicity, we compute a quadratic relation between Luminosity and the average metallicity: 
\begin{equation}
\begin{split}
    \frac{L_{\rm halo}}{L_{\rm halo, \langle [Fe/H] \rangle = -1.5}} &=& 1.0 -0.9851\big(\mathrm{\langle [Fe/H] \rangle} +1.5\big)\\[-2.0ex] &&+0.2670\mathrm{\big(\langle [Fe/H] \rangle+1.5\big)^2 } 
\end{split}
\end{equation}
Here, the halo luminosity can be adjusted from the fiducial estimate (assuming $\langle \mathrm{[Fe/H]} \rangle = -1.5$) using the above relation. Note that this equation is only valid for average metallicities in the range $-2.0 < \langle \mathrm{[Fe/H]} \rangle < -1.0$. For completeness, we also provide a conversion formula for the total stellar halo mass assuming a Kroupa IMF. Note the relation is not identical to the luminosity conversion (modulo a factor conversion) as the stellar-mass-to-light ratio depends on metallicity. For example, for a Kroupa IMF (and assuming old ages) the stellar-mass-to-light ratio is 1.6(1.4) for $\langle \mathrm{[Fe/H]} \rangle = -1.0(-2.0)$.
\begin{equation}
\begin{split}
    \frac{M^\star_{\rm halo}}{M^\star_{\rm halo, \langle [Fe/H] \rangle = -1.5}} &=& 1.0 -0.9104\big(\mathrm{\langle [Fe/H] \rangle}+1.5 \big)\\[-2.5ex] &&+0.2473\big(\mathrm{\langle [Fe/H] \rangle} +1.5\big)^2
\end{split}
\end{equation}
\cite{conroy19} recently reported that the average stellar halo metallicity is higher than previously thought, with $\langle \mathrm{[Fe/H]}\rangle =-1.2$. If we use this increased metallicity in the formula given above then the derived stellar halo mass is $M^\star_{\rm halo} = 1.05 \times 10^9M_\odot$, i.e. 25 \% lower than our fiducial estimate (see following subsection for further discussion). 

Finally, our stellar halo mass (and luminosity) estimate is also dependent on the suite of isochrones used in the analysis, as the predictions for the RGB can vary between different stellar population models \citep[see e.g.][]{hidalgo18}. If we repeat our analysis (assuming our fiducial density profile, MDF, IMF assumptions) with the MIST \citep{choi16} or BaSTI \citep{hidalgo18} models, we find (total) stellar masses of $M^\star_{\rm halo} = 0.85 \times 10^{9}M_\odot$ and $M^\star_{\rm halo} = 1.1 \times 10^{9}M_\odot$, respectively. These masses are slightly lower than our fiducial results (based on the PARSEC isochrones), but still consistent within the uncertainties. The complexities of modeling the RGB in isochrone libraries is beyond the scope of this paper, but this, in addition to the systematic effects mentioned above, is an important consideration for stellar halo mass estimates, and will need close attention in future work to achieve both precise and accurate measures.

\begin{figure*}
 \begin{minipage}{0.45\linewidth}
        \centering
        \includegraphics[width=8cm,angle=0]{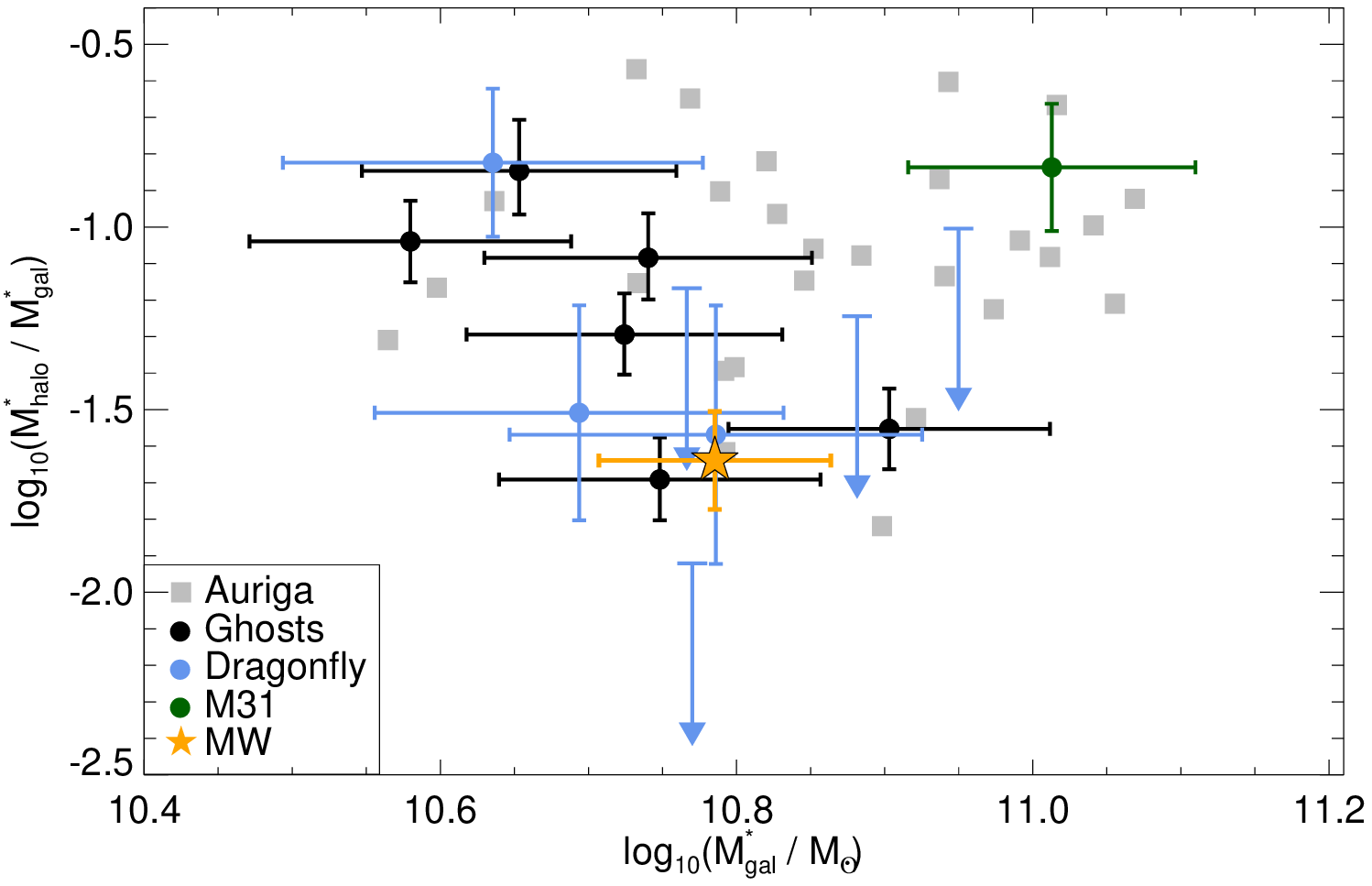}
  \end{minipage}
  \begin{minipage}{0.45\linewidth}
        \centering
        \includegraphics[width=8cm,angle=0]{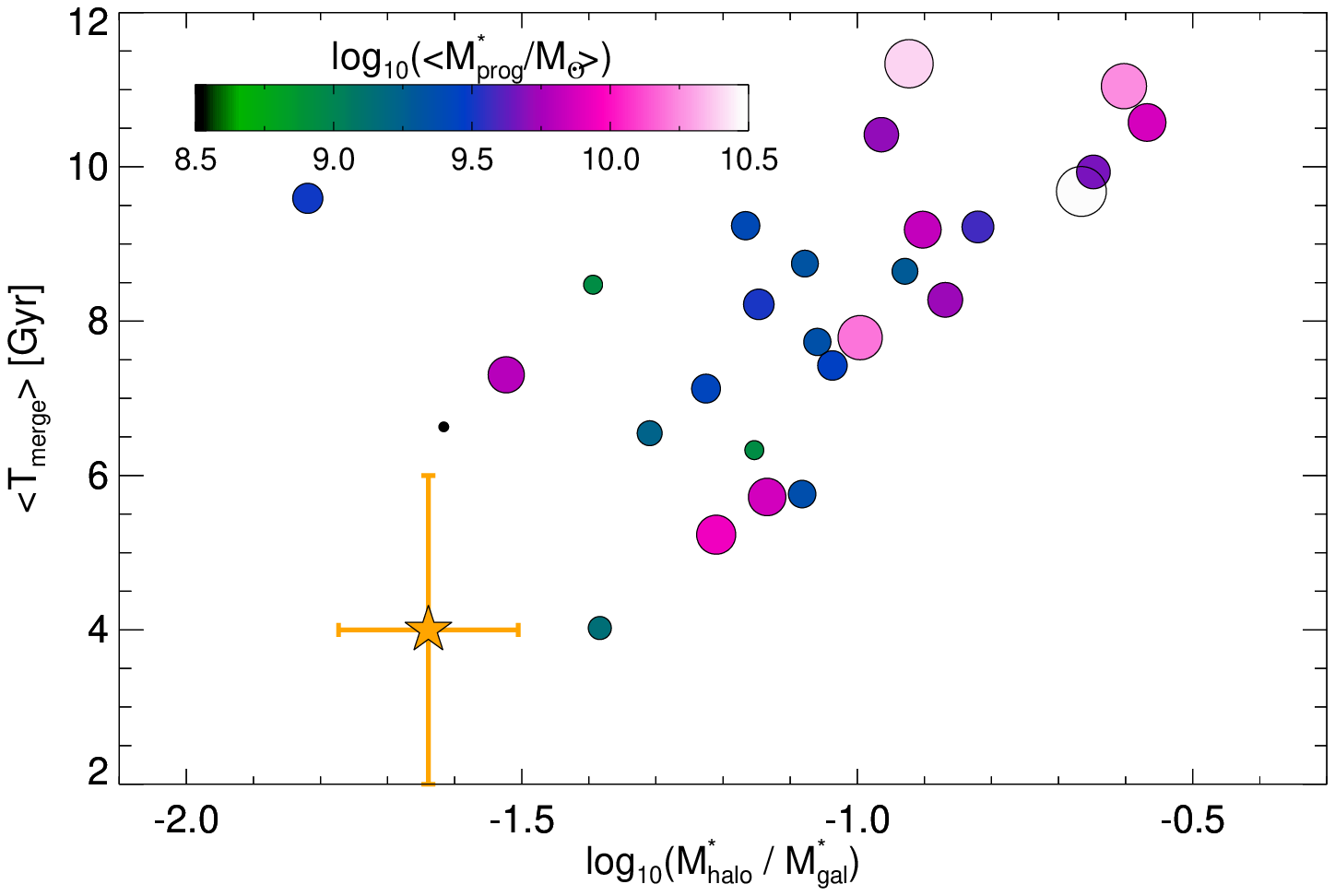}
  \end{minipage}
\caption[]{\textit{Left:} The stellar halo mass fraction
  ($M^\star_{\rm halo}/M^\star_{\rm gal}$) as a function of galaxy mass. We
  show the simulated Auriga galaxies, and observational estimates from
  Ghosts \citep{harmsen17}, Dragonfly \citep{merritt16}, and M31
  \citep{sick15, harmsen17}. The yellow star indicates our measure for
  the Milky Way assuming a Kroupa IMF (note we use the Galaxy mass
  from \citealt{licquia15}). \textit{Right:} The average merger time
  of Milky Way halo progenitors against stellar halo mass fraction for
  the \textsc{Auriga} haloes. The points are coloured (and scaled) according to
  the average progenitor mass. The Milky Way is indicated with the
  orange star.}
\label{fig:mhalo}
\end{figure*}

\subsection{Tension between stellar halo mass and metallicity?}
Dwarf galaxies follow a fairly tight ($\sim 0.2$ dex scatter) stellar mass-metallicity relation \citep{kirby13}. Following the relation derived by \cite{kirby13} based on local group galaxies, dwarfs with masses in the range $0.5-1 \times 10^9M_\odot$ have average metallicities of $\langle \mathrm{[Fe/H]} \rangle \sim -0.9$ to $-0.8$ dex. The average metallicity of halo stars is $\langle \mathrm{[Fe/H]} \rangle \sim -1.5$ \citep{an13, zuo17}, which is seemingly at odds with a stellar halo mass of $\sim 10^9M_\odot$ dominated by one massive progenitor. However, this simple exercise ignores two important factors: (1) we are using the $z=0$ mass-metallicity relation, and the stellar halo was built up in the past, and (2) the inner halo, within $\sim 20$ kpc is likely dominated by a massive progenitor, but the outer parts are likely biased towards lower mass contributors \citep{deason18, lancaster19}. 

\cite{deason16} use cosmological N-body simulations to explore the
relation between accreted stellar mass and metallicity. They used
empirical stellar mass-halo mass relations, and redshift dependent
stellar mass-metallicity relations, to map accreted dark matter
subhaloes to stellar halo progenitors. In their Figure 7 they show
that the relation between the average metallicity of the accreted
stellar material and the typical destroyed dwarf mass. For progenitors
of $M^\star \sim 0.5-1 \times 10^9M_\odot$, the average metallicity
varies between $\langle [\mathrm{Fe/H}] \rangle \sim$ -1.0 to
-1.5. The lower metallicities are only obtained when the progenitor is
destroyed at very early times, when the average metallicity of the
dwarf galaxies (at fixed mass) is lower \citep{ma16} (see also Fattahi
et al. in preparation). Thus, in order to reconcile the stellar halo
metallicity with a massive progenitor (and hence relatively massive
stellar halo), this event must have occurred $\gtrsim$ 10 Gyr ago. This
is exactly the scenario that has been proposed in the \gaia\--Sausage
or \gaia\--Enceladus discovery papers: an \textit{ancient}, massive
accretion event \citep{belokurov18, haywood18,helmi18}. 

Recently, \cite{conroy19} suggested that the average stellar halo metallicity should be revised upwards to $\langle [\mathrm{Fe/H}] \rangle = -1.2$. In this case, as discussed in the previous section, our estimated stellar halo mass is slightly lower ($\sim 1.0 \times 10^9M_\odot$ rather than $1.4 \times 10^9M_\odot$). The argument above --- that this metallicity-stellar halo mass combination favours an ancient accretion event --- still holds, but the disparity with the $z=0$ stellar mass-metallicity relation is less severe.

Finally, as mentioned earlier, although the bulk of the (inner) stellar halo mass may be contributed by the \gaia\--Sausage, there is still a sprinkling of lower mass, $\sim 10^7-10^8M_\odot$ progenitors (e.g. Sgr, Sequoia), with lower average metallicities, that also contribute to the total stellar halo mass (and average metallicity). Moreover, there could also be a contribution from \textit{in-situ} halo stars to the total stellar halo mass. We discussed in Section \ref{sec:rgb} that the Galaxia + N-body models do not include \textit{in-situ} halo stars. \cite{belokurov19} recently showed evidence for an \textit{in-situ} halo population, dubbed ``Splash'', in the inner Milky Way halo (see also \citealt{Gallart2019}). The Splash is kinematically hot and has chemical and kinematic features that are intermediate between halo and thick disc populations. However, importantly, \cite{belokurov19} show that the Splash is confined to the inner halo. Indeed, they find at heights of $|z| \sim 10$ kpc the Splash drops to a meagre 5\% of the halo density. As our analysis is mainly concerned with distant stars ($D \sim 5-100$ kpc) at high Galactic latitude ($|b| > 30^\circ$), we do not expect our halo mass estimate to be contaminated by more than 5\% from Splash stars. However, we do caution that the cosmological simulations do predict a signficant amount of distant \textit{in-situ} halo material \citep[see e.g.][]{monachesi19}. If this is true in the Milky Way, then our total stellar halo mass estimate is a combination of accreted halo stars and any \textit{in-situ} material that manages to make it out to significant distances in the halo.

\subsection{The Milky Way in context}

At fixed galaxy (or halo) mass, the stellar halo mass can vary significantly: this has been seen both in simulations and observations \citep[e.g.][]{pillepich14, merritt16, elias18, monachesi19}. Thus, the stellar halo mass is intimately linked to the assembly history of the halo. For example, if a halo is dominated by one accretion event, then the stellar halo mass will reflect the mass of this progenitor (see e.g. \citealt{deason16, dsouza18}.)

In the left panel of Fig. \ref{fig:mhalo} we show the ratio of stellar
masses of the accreted (halo) and the galaxy (host) populations
against the galaxy's stellar mass. Here, we show the values from the
$N=30$ \textsc{Auriga} simulations in grey. This is a suite of high
resolution cosmological hydrodynamic simulations of Milky Way mass
haloes (see \citealt{grand17} for more details). The stellar halo
masses we show here only include the ``accreted" stellar halo mass. As
noted by \cite{monachesi19}, the stellar halo masses in
\textsc{Auriga} are significantly overestimated if we do not excise
the halo stars born \textit{in-situ}. We also show observational
measurements in the left panel from the Ghosts (black filled circles,
\citealt{harmsen17}) and Dragonfly (pale blue filled circles,
\citealt{merritt16}) surveys. Our estimate for the Milky Way is shown
with the orange star symbol (assuming a Kroupa IMF). Here, we use
total Galactic stellar mass derived in \cite{licquia15}. Even though
our stellar halo mass is larger than some previous estimates in the
literature, the Milky Way stellar halo mass fraction is relatively low
compared to both external galaxies and the \textsc{Auriga}
simulations.

In the right hand panel we use the \textsc{Auriga} simulations to show
$M^\star_{\rm halo} / M^\star_{\rm gal}$ against the average merger time of
the destroyed dwarf galaxies that build-up the stellar halo ($\langle T_{\rm merge} \rangle$: computed by averaging over all star particles within 100 kpc). The
points are coloured (and scaled) according to the average progenitor
mass. Note that the quantities in Fig. \ref{fig:mhalo} (e.g. merger
times, accreted stellar mass) for the \textsc{Auriga} simulations are
derived in the works by \cite{fattahi19} and \cite{monachesi19}. There
is a clear trend between \textit{the epoch} of a dwarf accretion and
the fraction of stellar mass in the halo: earlier accretion events
lead to a lower fraction of stars in the halo (see also
\citealt{elias18}). Early mergers truncate the star-formation activity
in the progenitor dwarfs, while the dwarfs accreted later were able to
continue to form stars. Note that haloes with a large number of progenitors (e.g. Halo-17, blue point in top-left of right-hand panel) do not follow this trend as closely as the ``average'' progenitor mass and merger time are more ill-defined. For illustration, we indicate the Milky Way
with the orange star. Here, we have assumed the typical merger time
for the \gaia\--Sausage is $10 \pm 2$ Gyr ago (4 Gyr since the Big
Bang). Even though most haloes in \textsc{Auriga} experience more
recent accretion events, it is clear that an ancient merger event,
with little activity after said-event, can adequately explain the
stellar halo mass fraction of the Milky Way. 

In summary, our estimate stellar halo mass supports a scenario whereby
the Milky Way experienced an early ($\sim 10$ Gyr ago), massive
($M^\star \sim 0.5-1.0 \times 10^9M_\odot$) merger event, and had only
relatively minor mergers thereafter\footnote{At least until the LMC is
  digested, see \cite{cautun19}.}.

\section{Conclusions}
\label{sec:conc}
In this work we have used counts of RGB stars from GDR2 to estimate
the total stellar luminosity of the Milky Way's halo. Using slices in
colour, magnitude and position on the sky, we decompose the disc and
halo RGB populations using 2-dimensional Gaussian fits to the proper
motion distributions. The resulting counts of halo stars are converted
into a stellar luminosity using a suite of (weighted) PARSEC isochrones. Our
analysis is tested and calibrated on the \textit{Galaxia} model, using
the BJ05 N-body models for the stellar halo component. Our main
results are summarised as follows:

\begin{itemize}

\item In the majority (70\%) of bins in magnitude, colour and area on
  the sky we are able to recover the true number of halo RGB stars to
  $\lesssim 30\%$. Tests with the \textit{Galaxia}+BJ05 models show
  that we are able to recover the true total luminosity to within 25
  \% if the metallicity distribution and density profile of the halo stars are known. This confidence interval takes into account realistic systematic
  uncertainties, such as the presence of substructure and non-Gaussian
  proper motion distributions.

\item After applying our method to GDR2, and assuming an Einasto density profile \citep{deason11} and MDF with $\langle \mathrm{[Fe/H]} \rangle = -1.5$ for the stellar halo, we find a total luminosity of $L_{\rm halo} = 7.9 \pm 2.0 \times 10^8L_\odot$ excluding Sgr, and $L_{\rm halo} = 9.4 \pm 2.4 \times 10^8L_\odot$ including Sgr. The difference when Sgr is included or excluded gives a rough estimate of the total luminosity of the Sgr progenitor: $L_{\rm Sgr} \sim 1.5 \times 10^8 L_\odot$, in good agreement with the value derived by \cite{ostholt12}.

\item We explore additional systematic uncertainties from our adopted MDF and density profile for the halo. In particular, we find the metallicity strongly influences the derived luminosity, and we provide an approximate conversion formula to infer luminosity (and mass) for a different MDF. Moreover, differences in the literature regarding the halo density profile leads to an additional systematic uncertainty of $\sim 30\%$ in our derived luminosity and mass.

\item Assuming a stellar mass-to-light ratio appropriate for a Kroupa IMF ($M^\star/L = 1.5$), and our fiducial halo density profile and MDF, we estimate a stellar halo mass of $M^\star_{\rm halo} = 1.4 \pm 0.4\times 10^9 M_\odot$. This mass is larger than estimates in the literature using different stellar halo tracers (main sequence turn-off stars, blue horizontal branch stars), and different methods. However, a mass of $\sim 10^9M_\odot$ confirms the emerging picture that the (inner) stellar halo is dominated by one massive dwarf progenitor.

\item We show that haloes in the \textsc{Auriga} simulations that have similar stellar halo mass fractions ($M^\star_{\rm halo}/M^\star_{\rm gal} \sim 0.02 $) to the Milky Way are likely formed by ancient ($\sim 10$ Gyr) mergers. Indeed, the relatively low stellar halo mass fraction, and average metallicity of the stellar halo can only be reconciled with a massive progenitor if this was a very early merger event.
\end{itemize}

\section*{Acknowledgements}
We thank an anonymous referee for providing a thorough and insightful report, which greatly improved the quality of this manuscript.

AD thanks Azi Fattahi for help with simulation data, and Russell Smith for useful discussions regarding stellar population models. We thank the Auriga team for allowing us to use their data in Figure 12.

AD is supported by a Royal Society University Research Fellowship. AD also
acknowledges the support from the STFC grant ST/P000541/1. JLS acknowledge the support of the Leverhulme and Newton Trusts.

This work has made use of data from the European Space Agency (ESA) mission
\gaia\ (\url{https://www.cosmos.esa.int/gaia}), processed by the \gaia\
Data Processing and Analysis Consortium (DPAC,
\url{https://www.cosmos.esa.int/web/gaia/dpac/consortium}). Funding for the DPAC
has been provided by national institutions, in particular the institutions
participating in the \gaia\ Multilateral Agreement.

We acknowledge the Gaia Project Scientist Support Team and the DPAC. 

This work
used the DiRAC Data Centric system at Durham University, operated by
ICC on behalf of the STFC DiRAC HPC Facility (www.dirac.ac.uk). This
equipment was funded by BIS National E-infrastructure capital grant
ST/K00042X/1, STFC capital grant ST/H008519/1, and STFC DiRAC
Operations grant ST/K003267/1 and Durham University. DiRAC is part of
the National E-Infrastructure. 

AD thanks the staff at the Durham University Day Nursery who play a key role in enabling research like this to happen.

\bibliographystyle{mnras}
\bibliography{mybib}

\bsp	
\label{lastpage}
\end{document}